\documentclass[12pt]{iopart}
\usepackage{axodraw,latexsym,amssymb}
\usepackage{iopams} 
\usepackage{graphicx,epsf}

\newcommand{\BlackHat}{{\sc BlackHat}}
\newcommand{\SHERPA}{{\sc SHERPA}}
\newcommand{\Helac}{{\sc Helac}}
\newcommand{\CutTools}{{\sc CutTools}}

\newcommand{\MCFM}{{\sc MCFM}}

\newif\ifdraft
\drafttrue
\newif\ifpreprint
\preprinttrue

\def\fig#1{fig.~{\ref{#1}}}
\def\Fig#1{Fig.~{\ref{#1}}}

\def\sect#1{section~{\ref{#1}}}
\def\eqn#1{eq.~(\ref{#1})}
\def\Eqn#1{Equation~(\ref{#1})}

\def\pb{\bar p}

\def\e{\epsilon}

\def\Wjj{$W\,\!+2$}
\def\Wjjj{$W\,\!+3$}
\def\Wjjjj{$W\,\!+4$}

\def\Wjn{$W+n$}

\def\Zjj{$Z\,\!+2$}
\def\Zjjj{$Z\,\!+3$}
\def\Zjjjj{$Z\,\!+4$}

\def\Zjjjj{$Z\,\!+4$}

\def\eps{\epsilon}

\def\extraskip{\vskip 15mm}
\def\Ord{{\cal O}}
\def\A#1{{\cal A}_{#1}}

\def\tree{{\rm tree}}

\def\Tr{\mathop{\rm Tr}\nolimits}

\def\oneloop{{\rm 1\hbox{-}loop}}

\def\Gr{{\rm Gr}}
\def\L{\left(}\def\R{\right)}
\def\spa#1.#2{\left\langle#1\,#2\right\rangle}
\def\spab#1.#2.#3{\left\langle#1|#2|#3\right]}
\def\spaa#1.#2.#3{\left\langle#1|#2|#3\right\rangle}
\def\spb#1.#2{\left[#1\,#2\right]}
\def\spba#1.#2.#3{\left[#1|#2|#3\right\rangle}

\def\Ksl{{\s K}}
\def\ksl{\s{k}}
\def\tlambda{{\tilde\lambda}}
\def\ve{\epsilon}

\def\I{ {\cal I}}
\def\stack#1#2{\stackrel{#1}{\scriptscriptstyle #2}}

\def\NeqFour{{\cal N} = 4}
\def\NeqOne{{\cal N} = 1}

\def\cg{c_\Gamma}

\def\Shift#1#2{{[#1,#2\rangle}}

\def\inlimit^#1{\buildrel#1\over\longrightarrow}

\newbox\charbox
\newbox\slabox
\def\s#1{{      
        \setbox\charbox=\hbox{$#1$}
        \setbox\slabox=\hbox{$/$}
        \dimen\charbox=\ht\slabox
        \advance\dimen\charbox by -\dp\slabox
        \advance\dimen\charbox by -\ht\charbox
        \advance\dimen\charbox by \dp\charbox
        \divide\dimen\charbox by 2
        \raise-\dimen\charbox\hbox to \wd\charbox{\hss/\hss}
        \llap{$#1$}
}}

\def\nuornub{{}^{\raise1.3pt\hbox{\ttiny(}}\hskip -0.2pt\overline{\kern 
-0.5pt \nu \kern -0.4pt}\hskip 0.3pt{}^{\raise1.3pt\hbox{\ttiny)}}}

\begin{document}

\hbox{
\hskip 10.cm UCLA-11-TEP-120, NSF-KITP-11-175
}

\title{Susy Theories and QCD: Numerical Approaches.}

\author{Harald Ita}

\address{
\centerline{
Department of Physics and Astronomy, UCLA,}
\centerline{ 
Los Angeles, CA 90095-1547, USA} 
}

\ead{ita@physics.ucla.edu}

\begin{abstract}
We review on-shell and unitarity methods and discuss their application to
precision predictions for LHC physics.  Being universal and numerically robust,
these methods are straight-forward to automate for next-to-leading-order
computations within Standard Model and beyond.  Several state-of-the-art
results including studies of $W$/$Z$+3-jet and $W$+4-jet production have
explicitly demonstrated the effectiveness of the unitarity method for
describing multi-parton scattering. Here we review central ideas needed to
obtain efficient numerical implementations. This includes on-shell loop-level
recursions, the unitarity method, color management and further refined tricks.

This article is an invited review for a special issue of Journal of Physics A
devoted to ``Scattering Amplitudes in Gauge Theories''.  
\end{abstract}

\maketitle

\section{Introduction}

The physics program at the Large Hadron Collider (LHC) relies heavily on the
ever increasing theoretical control over modeling high-energy proton
collisions.  The detailed theoretical understanding not only increases the
reach in new physics and particle searches, but also allows to study the
fundamental dynamics and properties of particles.  Formulating new observables
for addressing specific physics questions is a typical task which relies on
quantitative reliable theoretical input. 
In the long run, high statistics measurements combined with precision
prediction from theory will allow systematic probes of fundamental particle
theory at ever deeper levels.  

The theory of Quantum Chromodynamics (QCD) concisely describes the collisions
of protons, however, the dominant dynamics differ depending on observables and
the regions of
phase-space~\cite{SalamElements,JPG,ESW,ManganoReview,DixonReview}.  Here we
have in mind proton collisions with large momentum transfer that are typical
for the production of heavy particles. These include the Higgs, top, new
particles within theories of supersymmetry as well as more conventional
Standard Model processes at large scattering angles. To a good approximation
such scattering processes factorize into the long-distance dynamics of quarks
and gluons within protons, short distance hard interactions between these
partons and, finally,  the formation of hadrons and observable jets from the
emerging partonic states and remaining proton fragments.  Monte Carlo event
generators deal with all aspects of the multi-layered simulation of the proton
collisions.  For some purposes, i.e.~sufficiently inclusive observables, it is
accurate to use a simpler, purely partonic description of events.  To this end
one combines final-state partons into observable jets, consistently ignoring
corrections from showering and hadronization.  Numerical methods are commonly
used for the evaluation of differential cross sections, being well suited for
comparisons to experimental data.  

The hard scattering process is a central stage in the simulation of proton
collisions.  It is described through scattering amplitudes which are accessible
through first principle computations in quantum field theory.  The complexity
of scattering process allows only a perturbative approach, based on the
expansion in the strong coupling $\alpha_S(\mu_R^2)$.  A first step towards a
theoretical understanding of QCD is the evaluation of cross sections at leading
order (LO) in the strong coupling $\alpha_S(\mu_R^2)$.  Many
tools~\cite{LOPrograms,HELAC,Amegic,COMIX,OMEGA} are available to generate
predictions at leading order.  Some of the methods applied incorporate
higher-multiplicity leading-order matrix elements into parton-showering
programs~\cite{PYTHIAetc,Sherpa}, using matching (or merging)
procedures~\cite{Matching,MLMSMPR}.  

The truncation of the perturbative expansion introduces an explicit dependence
on the unphysical renormalization scale $\mu_R$ leading to a theoretical
uncertainty.  QCD cross sections can have strong sensitivity on higher-order
corrections, motivating the challenging quest for perturbative corrections.
Next-to-leading (NLO) order predictions significantly reduce renormalization-
and factorization-scale dependence -- a feature that becomes more significant
with increasing jet multiplicity (see e.g.~\cite{LesHouches}).  In addition,
NLO corrections take into account further physics effects including initial
state radiation, parton merging into jets and a more complete account of
partonic production channels.  Fixed-order results at NLO can also be matched
to parton showers~\cite{MCNLO} with the prospect of complete event generation
at next-to-leading-order in the strong coupling.  The value of a first
principle understanding of scattering processes in addition to the increased
quantitative control motivates the quest for cross sections at NLO and
sometimes beyond this~\cite{HNNLO,NNLOVrap}. 

In this contribution we will describe loop-level
on-shell~\cite{Bootstrap,CoeffRecursion,BlackHatI} and unitarity
methods~\cite{UnitarityMethod,DdimUnitarity}. Our main focus will be on
generalized unitarity~\cite{Zqqgg,BCFUnitarity,Darren} and its extensions for
the numerical computation~\cite{OPP,EGK,BlackHatI,GKM} of hard scattering
matrix elements.  In addition, we will discuss numerical on-shell loop-level
recursions~\cite{BlackHatI}.  The central strategy of these approaches is to
make maximal use of universal physical principles and mathematical structures
in order to describe complex multi-particle processes including quantum
corrections.  Numerical algorithms based on unitarity and on-shell methods are
efficient, numerically stable and can be automated for generic scattering
processes. The rapid recent developments in this field have already led to many
new studies of complex scattering processes at the
LHC~\cite{PRLW3BH,EMZW3Tev,W3jDistributions,MStop,TeVZ,OPPNLO,W4jets,OtherUnitarityNLO,Madloop}.
Several further implementations of unitarity methods have been
reported~\cite{OtherLargeN,GKW,SAMURAI,Madloop,BBU}.  For related analytic
approaches see the chapter in this review by Britto~\cite{BReview}.

For processes at the LHC with complicated final states, computations can be
very challenging. In principle, one can use the traditional Feynman diagram
representation of amplitudes for numerical evaluation. However, even at leading
order, the number of Feynman diagrams grows more than factorial as the number
of final-states increases. Computation times scale accordingly and refined
approaches are needed. 

At leading order recursive approaches allow to reduce the growth in complexity.
Two central strategies are commonly used for numerical computations: Off-shell
recursion relations~\cite{BGrecursion,ALPHA,CHelSampling}, based on
Dyson-Schwinger equations, optimise the reuse of recurring groups of Feynman
graphs.  In contrast, on-shell recursions~\cite{BCFW,CSW} take advantage of the
remarkable simplicity of the physical scattering amplitudes (see
e.g.~\cite{TreeAmplitudes}).  The simplicity arises in part from symmetry
properties of tree
amplitudes~\cite{TreeSymmetries,Integrability,amplitudestructure} that are
present in QCD-like theories; see the chapter in this review by Brandhuber,
Spence and Travaglini~\cite{BSTReview}.  For most practical
purposes the efficiency of the two approaches is comparable, depending on the
explicit realization of the algorithms. 

At next-to-leading-order additional challenges arise, in particular, for the
virtual corrections due to the loop-momentum integration.  NLO cross sections
are built from several ingredients: virtual corrections, computed from the
interference of tree-level and one-loop amplitudes; real-emission corrections;
and a mechanism for isolating and integrating the infrared singularities in the
latter. Automated approaches~\cite{GKDipoles,OtherDipoles,FKS} to deal with
these issues are available based on subtraction
methods~\cite{FKSTheory,CS,Antenna}. Recursive methods are effective for
computations of real-emission corrections. Such methods, however, are not
directly applicable to virtual corrections.  Traditional methods evaluate the
loop integrals of Feynman diagrams (see e.g.~\cite{EGZ,BDDP,TraditionalFD}),
and have to overcome two central challenges: growth of the number of Feynman
diagram expressions and the evaluation of tensorial loop integrals, while
maintaining gauge invariance.  Means to deal with tensor integral
reductions~\cite{Denner,IntegralReduction} as well as strategies to recycle
substructures have been shown to reduce complexity inherent in Feynman diagram
approaches~\cite{ElectroWeak4f,BDDP}. For a more complete discussion of
important NLO computations along these lines see ref.~\cite{LHSummary}.

The unitarity method~\cite{UnitarityMethod}, in contrast, constructs loop
amplitudes from on-shell tree amplitudes; gauge invariance is built in and
maintained throughout computations.  In addition, complexity arising from large
numbers of Feynman diagrams is avoided by recursive methods for tree
evaluations.  Similarly, loop-level
recursions~\cite{Bootstrap,CoeffRecursion,Genhel,BlackHatI} construct
amplitudes efficiently using purely on-shell lower-point input.  Effective
numerically stable implementations of these on-shell methods have been
demonstrated by various
groups~\cite{BlackHatI,GZ,OPPNLO,GKW,OtherLargeN,SAMURAI,Madloop,BBU}.  Beyond
this, unitarity approaches have already been applied to state-of-the-art NLO
computations~\cite{W4jets,PRLW3BH,EMZW3Tev,W3jDistributions,MZ3j,TeVZ,OtherUnitarityNLO}.

In more detail, numerical unitarity based approaches use a combination of
methods.  Scattering amplitudes are naturally split into two parts; a part with
logarithmic dependence on kinematic invariants and a rational remainder.
Typically, unitarity approaches in strictly four dimensions are used for the
computation of the logarithmic parts, although, on-shell
recursions~\cite{CoeffRecursion} may as well be applied in certain cases.  At
present, there are three main choices for computing the rational remainder
within a process-nonspecific numerical program: on-shell
recursion~\cite{BlackHatI}, $D$-dimensional unitarity~\cite{DdimUnitarity,GKM},
and a refined Feynman-diagram approach~\cite{HPP,Madloop}.  We will discuss
here numerical unitarity approaches in four and $D$ dimensions following
refs.~\cite{BlackHatI,GKM} as well as numerical loop-level on-shell
methods~\cite{BlackHatI}.

Several recent developments allow us to use a purely numerical approach at loop
level.  A key tool is generalized unitarity~\cite{Zqqgg,BCFUnitarity} which
imposes multiple unitarity cuts (more then two propagators) and gives a refined
system of consistency relations that is easier to solve.  In addition,
generalized cuts allow for a hierarchical approach; computing coefficients of
four-point integrals, in turn three-point and, finally, two-point etc.
integrals from successively cutting four, three and two propagators. (For the
related maximal cut technique for multi-loop amplitudes see~\cite{CJReview}.)
Analogous approaches may be applied in $D$ dimensions~\cite{DdimUnitarity} (see
also~\cite{DdimUnitarityRecent}), where higher point integrals have to be
considered.  Further simplifications arise from working directly with the loop
integrand as opposed to the integrated loop amplitude.  The unitarity method
then turns into a purely algebraic approach. 
The starting point for this approach is a generic representation of the loop
integrand~\cite{OPP}, whose free, process dependent parameters are to be
determined by unitarity relations.  A particularly useful parametrisation of
the loop integrand was given by~\cite{OPP} (see also ref.~\cite{EGK}) and
extended in~\cite{GKM} to $D$ dimensions.  Importantly, despite the
restrictions imposed by on-shell conditions in unitarity cuts, sufficient
freedom remains in loop-momentum parametrization in order to uniquely determine
all integral coefficients for renormalizable theories and beyond.  

Modern on-shell and unitarity methods may be set up to take advantage of
refined physical properties and formal structures of scattering amplitudes.  We
will discuss the uses of structures like spinor
helicity~\cite{SpinorHelicity,TreeReview}, analyticity properties, color
decomposition~\cite{BGColor,BKColor,Neq4Oneloop} and supersymmetry
properties~\cite{TwoQuarkThreeGluon,Zqqgg} in order to make computations
efficient.  The scattering amplitudes are then decomposed into a fine set of
gauge invariant pieces (primitive amplitudes), which are computed individually
and eventually assembled into the full matrix element.  This approach leads to
excellent numerical stability and can be further
exploited~\cite{W3jDistributions} for caching and efficiency gains through
importance sampling (as used in color-expansions).  In addition to aiming for
efficiency it can be helpful to use methods which are easy to automate within
existing frameworks~\cite{HPP,Madloop} or fulfill further computational
constraints~\cite{GKW}.

Furthermore, a numerical approach to amplitudes requires attention to numerical
instabilities induced by round-off error. A natural way to deal with round-off
errors is to setup a rescue system which monitors numerical precision and
invokes an alternative computational approach when checks fail. A convenient
rescue strategy~\cite{CutTools,BlackHatI} for unitarity based approaches is the
use of higher precision arithmetic.  The advantage of a fine split-up of loop
amplitudes into gauge invariant subparts is that one can setup a very targeted
and thus efficient rescue system.

The present chapter of this review is organized as follows. In
\sect{ExampleW4j} we discuss 
a representative example for next-to-leading-order multijet
computations at hadron colliders pointing out the importance of on-shell and
unitarity methods. In \sect{StructureMatrixElements} we discuss the key
properties of in matrix elements that can be exploited for the computation of
loop amplitudes.  Finally, numerical unitarity approaches will be discussed
in~\sect{UnitaritySection} and loop-level recursions
in~\sect{OnShellRecursionSection}. We end with conclusions and an outlook.


\extraskip

\section{NLO Predictions for Hadron Colliders}
\label{ExampleW4j}

As an example, to point out the key features of NLO QCD predictions we will
focus on processes of massive vector-boson production in association with jets.
In particular, we focus on recent progress due to the use of numerical
unitarity approaches.

The production of massive vector bosons in association with jets at hadron
colliders has been the subject of theoretical studies for over three
decades~\cite{EarlyVplus1,Wplus2ME,EarlyWplus2,EarlyWplus2MP,Vplus1NLO,
Vplus1NLOAR,Vecbos,CampbellEllisZW}.  These theoretical studies have for
example played an important role in the discovery of the top
quark~\cite{TopQuarkDiscovery}.  The one-loop matrix elements for \Wjj-jet and
\Zjj-jet production were determined~\cite{Zqqgg} via the unitarity
method~\cite{UnitarityMethod} (see also ref.~\cite{OtherZpppp}), and
incorporated into the parton-level \MCFM~\cite{MCFM} program.  Studies of $W$
production in association with heavy quarks have also been
performed~\cite{WcNLO,WbjNLO,WZMassiveProgNLO,Wplus2MassiveNLO,WbNLO,BCEWb}.  

Beyond this, the numerical unitarity approach allowed to include additional
final-state objects. Studies of \Wjjj{}-jet production can be found in
\cite{PRLW3BH,EMZW3Tev,W3jDistributions,MZ3j} and \Zjjj{}-jet in~\cite{TeVZ}.
The state-of-the-art in perturbative QCD for hadron colliders are currently
parton-level next-to-leading order computations with up to five final-state
objects.  The first and only such process to be computed so far is \Wjjjj{}-jet
production~\cite{W4jets}.  More generally, several important QCD processes with
four final-state objects have been computed to
date~\cite{OtherNLO,OPPNLO,ee5jNLO,OtherUnitarityNLO}.  

Processes of $Z$- and $W$-boson production in association with jets have a
particularly rich phenomenology at the electro-weak symmetry breaking scale,
being important backgrounds to many searches for new physics and particles, for
Higgs physics, and will continue to be important to precision top-quark
measurements.  Decays of the massive vector-bosons into neutrinos mimic missing
energy signals and are of particular importance for supersymmetry
searches~\cite{SusySearch,ATLASSusy,CMSSusy}.  The clean leptonic decays of the
vector-bosons open a high resolution view of underlying QCD dynamics.
Inclusive production cross section provides valuable information about parton
distribution functions as well as fundamental parameters of the Standard Model.
The signal of vector-boson production in association with jets per se includes
physics of jet-production ratios~\cite{BerendsRatio} including comparative
studies of $W$, $Z$ and photon production.  Experimental studies of
vector-boson + jet production at the Tevatron were published by the CDF and D0
collaborations~\cite{WCDF,CDFZ,D0Z} as well as at the LHC by the ATLAS
collaboration~\cite{ATLASW}.

\subsection{Validation \& Prediction}
\label{Validation}

Before turning to the theoretical and technical issues it is useful to assess
how good the results are by comparing to experimental data.  The quantitative
impact of NLO corrections can be directly validated against data from Tevatron
$p\pb$ collisions.  
%
\begin{figure*}[tbh] \centerline{
	\includegraphics[clip,scale=0.385]{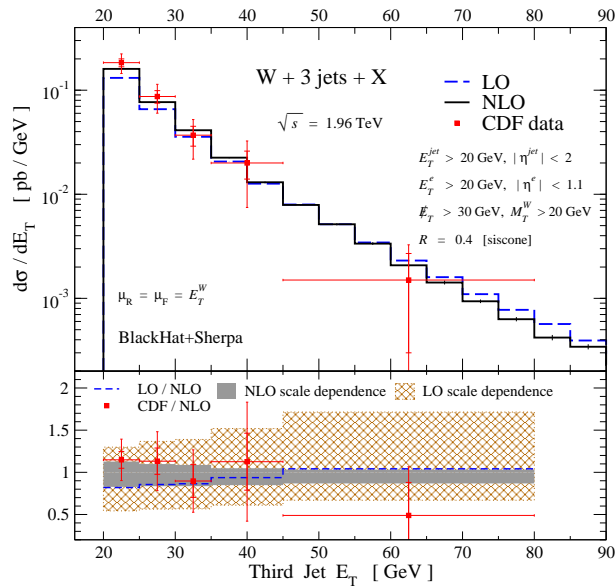}
	}

\caption{
The measured $E_T$ distribution of the softest observed jet
in inclusive \Wjjj-jet production,
compared to the NLO prediction~\cite{PRLW3BH,W3jDistributions}.  In the upper
panels the NLO distribution is the solid (black) histogram, and CDF data points
are the (red) points, whose inner and outer error bars, respectively, denote
the statistical and total uncertainties (excluding the luminosity error) on the
measurements.  The LO predictions are shown as dashed (blue) lines.  The lower
panel shows the distribution normalized to the full NLO prediction.  The
scale-dependence bands are shaded (gray) for NLO and cross-hatched (brown) for
LO.\\ 
{\it\tiny Reprinted \fig{W3TevatronFigure} with permission from~\cite{W3jDistributions} p.27. Copyright (2009) by the American Physical Society.} } 
\label{W3TevatronFigure} \end{figure*}
%
\Fig{W3TevatronFigure} compares the $E_T$ distribution of the third-most
energetic jet in CDF data~\cite{WCDF} to the NLO and LO predictions for
\Wjjj-jet production~\cite{PRLW3BH,W3jDistributions}. (For a similar analysis
using a leading color approximation see also~\cite{EMZW3Tev}.)  The upper
panels of \fig{W3TevatronFigure} show the distribution itself, while the lower
panels show the ratio of the LO value and of the data to the NLO result.
The NLO predictions match the data very well, and uniformly with well matching
slope.  The central values of the LO predictions, in contrast, have different
shapes from the data.  The change of shape between LO and NLO is attributed to
an imprecise scale in the coupling constant, that governs the production of the
softest observed jet, in the leading order computation which gets corrected
once loop effects are included as discussed in
refs.~\cite{BauerScale,EMZW3Tev}.

Scale-dependence bands indicate rough estimates of the theoretical error.
Those are obtained by varying the renormalization- and factorization scales by
factors of two around a central scale.  
For such scale variations, the dependence is of the order of $\pm40\%$
for \Wjjj-jet processes at LO.  
At NLO, the scale dependence shrinks to $\pm10\%$, and we
obtain a quantitatively reliable answer. A more detailed discussion can be found in
refs.~\cite{PRLW3BH,W3jDistributions}.

As another example, we consider predictions for the LHC.  For the
inclusive production of \Wjn{} jets, a basic quantity to examine is the $p_T$
distribution for the softest observed jet.  
%
\begin{figure*}[th]
	
	\centerline{
	\includegraphics[clip,scale=0.5]{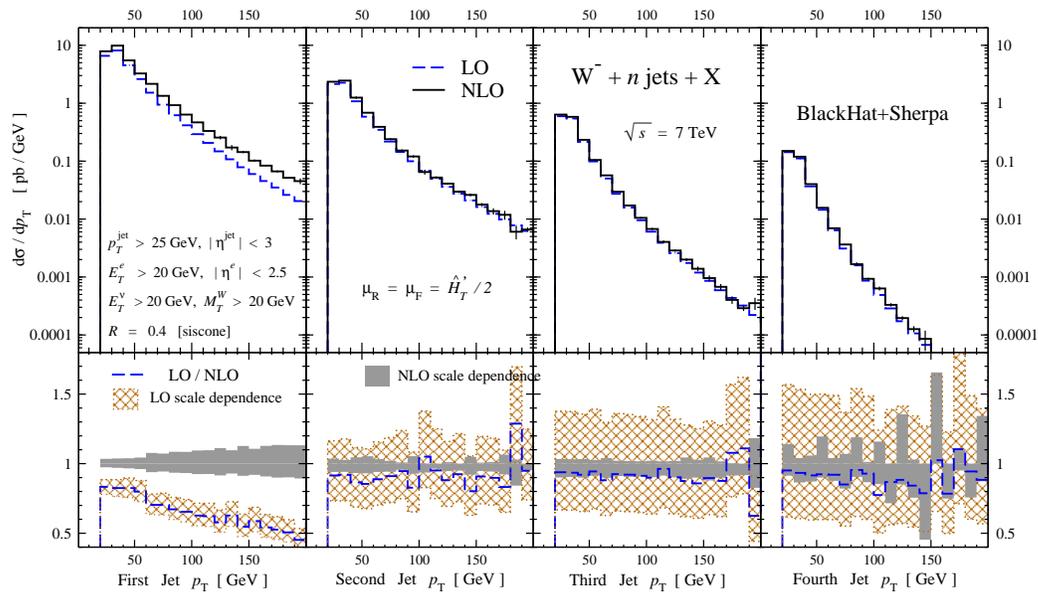}
	}
	\caption{A comparison of the $p_T$ distributions of the softest
	observable jets in $W^-$+n-jet ($n=1,2,3,4$) production, respectively.
	The setup describes the LHC proton-proton collisions at
	$\sqrt{s}=7$~TeV as published in~\cite{W4jets}.  In the upper panels
	the NLO distribution is the solid (black) histogram and the LO
	predictions are shown as dashed (blue) lines.  The lower panels show
	the LO distribution and LO and NLO scale-dependence bands normalized to
	the central NLO prediction.  
	} \label{W4ptFigure} \end{figure*}
%
\Fig{W4ptFigure} shows the $p_T$ distributions of the softest observed jet in
$W^−+n$-jet ($n=1,2,3,4$) production at LO and NLO, respectively.  For details
on our analysis setup we refer to~\cite{W4jets}.  The predictions are
normalized to the central NLO prediction in the lower panels.  Comparing the
$p_T$ distributions successively starting from $W^−$+1-jet production, we
observe the reduction in differential cross section of about a factor of
$\alpha_S$ from one panel to the next; each observed jet leads to an additional
power in the strong coupling.  At the same time the lower panels in
\fig{W4ptFigure} show an approximately linear increase in the LO scale
variation bands, up to about $\pm50\%$ for \Wjjjj{} jets. The scale variation
of the NLO result, displayed in the lower panels of \fig{W4ptFigure}, is
strongly reduced to about $\pm10\%$ for the present setup.

In summary, in the above examples the advantages of NLO computations over the
leading order appear through several quantitative improvements.  Firstly, NLO
predictions show a greatly reduced dependence on unphysical renormalization and
factorization-scales as compared to leading order.  The second improvement we
pointed out concerns the shapes of distributions.  Due to inclusion of
radiation from an additional parton as well as a more truthful description of the
scale dependence shapes of distributions are modeled better at NLO.

The above results validate our understanding of the \Wjn{}-jet
processes for typical Standard-Model cuts.  It will be interesting, and
necessary, to explore the size of corrections for observables and cuts used in
new-physics searches.  A related process that contributes an irreducible
background to certain missing energy signals of new physics is \Zjjjj-jet
production.  We expect that the current setup~\cite{W4jets} will allow us to
compute NLO corrections to \Zjjjj-jet production, as well as to other complex
processes, thereby providing an unprecedented level of theoretical precision
for such backgrounds at the LHC.

Parton-level NLO simulations of this kind are first principle predictions
whose outcome directly reflect properties of the underlying theory. 
Although NLO computations are more challenging, in general they yield results
with better reliability and agreement with measurements.

\subsection{Setup of Complete Computation}

The computation of differential distributions is the end product of combining
many important ingredients pulled together in a Monte Carlo program; these
include parton distribution functions and couplings, phase-space integration,
matrix elements, analysis framework etc.  Various tools are available to deal
with complete NLO computations. One such tool is \MCFM~\cite{MCFMweb}, which
contains an extensive library of analytic matrix elements for NLO computations.
Another approach (see~\cite{OPPNLO} and references) uses tools including
\Helac~\cite{HELAC} and \CutTools~\cite{CutTools} for a numerical
approach~\cite{OPP}.  Here we will describe another
setup~\cite{PRLW3BH,W3jDistributions,W4jets}  based on on-shell and unitarity
methods that was used for the computation of \Wjn{}-jet production in
\sect{Validation}.

In addition to LO components of Monte-Carlo programs, at NLO the computations
rely on further similarly important ingredients.  For the
\Wjn{}-jet production~\cite{PRLW3BH,W3jDistributions,W4jets} the real-emission
and dipole-subtraction terms~\cite{CS}, are provided by the \SHERPA{}
package~\cite{Sherpa}. \SHERPA{} was used to perform phase-space
integration.  \BlackHat{} was used to compute the real-emission tree amplitudes
for \Wjjjj jets using on-shell recursion relations~\cite{BCFW}, along with
efficient analytic forms extracted from ${\cal N}=4$ super-Yang-Mills
theory~\cite{DrummondTrees}.  
%
\begin{figure} 
	\begin{center} \centerline{ \epsfxsize 4
	truein\epsfbox{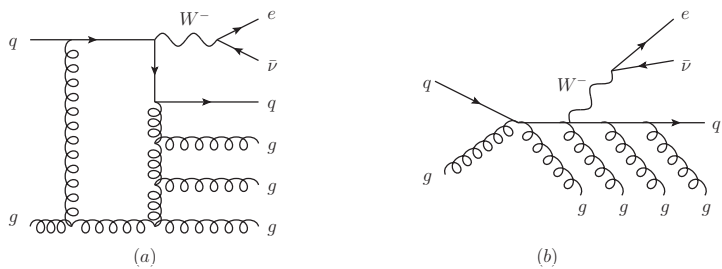} } \caption{ Representative
	diagrams of matrix elements for \Wjjjj-jet production: (a) the eight-point loop amplitudes
	$qg\rightarrow e\bar\nu q' ggg$, and (b) the nine-point tree-level
	amplitudes $qg\rightarrow e\bar\nu q' gggg$ needed for the real
	contribution.  The $e\bar\nu$ pair couples to the quarks via a W boson.
	}\label{FeynmanDiagrams} 
\end{center} \end{figure}
%

In terms of scattering amplitudes we need the input of up to eight-point
one-loop QCD amplitudes as well as up to nine-point tree-level QCD amplitudes;
example Feynman diagrams are shown in~\fig{FeynmanDiagrams}. The squared matrix
elements are summed over all initial and final-state partons, parton helicities and
color-orderings. For the present computation the $W$-boson is decayed into an
observable electron and a neutrino.  Amplitudes of this kind can be obtained
from QCD amplitudes; with the lepton pair replaced by a quark pair and the
$W$-boson exchange related to a gluon exchange.  Appropriate dressing with
coupling constant and propagator terms are needed. A recent analysis of high
multiplicity tree amplitudes of this kind can be found in~\cite{DrummondTrees}.

\section{Structure of One-loop Matrix Elements}
\label{StructureMatrixElements}

The evaluation time of matrix
elements is often dominating cross section computations, thus, emphasising the
importance of efficient numerical algorithms.  Beyond this, matrix elements are
objects of fundamental theoretical interest; new physics effects observable at
high energy colliders may originate in properties of matrix elements (see
e.g.~\cite{Wpol}).  

Matrix elements are functions of a large number of variables, which
characterize particles, polarization states, color quantum numbers, and
kinematics. To next-to-leading-order in the strong coupling the parton level
cross sections for $(N-2)$ resolved final-state objects, $pp\rightarrow (N-2)$,
depend on squared born matrix elements,
\begin{eqnarray}
\sum_{a_i,h_i}{|\A{n}^{\tree}(\{k_i,h_i,a_i\}|^2},\qquad\qquad
n=N,N+1\,,
\label{ColorHelicitySum}
\end{eqnarray}
as well as the interference terms,
\begin{eqnarray}
	\sum_{a_i,h_i}{\A{n}^{\tree\,*}(\{k_i,h_i,a_i\})
	\A{n}^\oneloop(\{k_i,h_i,a_i\})}\,+\,c.c.\,,\,\qquad n=N\,,
\end{eqnarray}
where $k_i$, $h_i$, and $a_i$ are respectively the momentum, helicity ($\pm$),
and color index of the $i$-th external gluon or quark. The shorthand $c.c.$
refers to the complex conjugate part that has to be included. The efficient
management of parton and helicity sums is important.  For simplicity, we will
consider scattering amplitudes involving quarks and gluons only. Much of the
methods can be carried over to more general particle spectra.

As inspired by analytic approaches (see e.g.\cite{OnShellReview}) we
disentangle degrees of freedom in order to arrive at a fine set of gauge
invariant objects.  To this end several
structures are used: Color decomposition into color ordered sub amplitudes
disentangles color information and kinematics.  Use of spinor helicity notation
aligns notation of kinematics and polarization vectors. Spinor
variables, in addition, lead to a natural way to work in complex momentum space.
This in turn allows to exploit analyticity properties of the basic color ordered scattering
amplitudes. Use of a standard basis of integral functions will allow a further
fine split-up of the loop amplitude into integral functions and their integral
coefficients.

Several features motivate us to disentangle matrix elements into a fine set of
primitive amplitudes.  First of all, do these have cleaner physical and
analytic properties than the full matrix elements, as will be discussed in
detail below.  This can be exploited for the construction of computational
algorithms, as will be discussed in detail below. For numerical approaches, a
detailed understanding of physical properties (e.g.  IR/UV pole structure of
primitive amplitudes, or, tensor rank of integrals) allows one to monitor
precision and stability of the computation.  Furthermore, caching systems built
on primitive objects (here, amplitudes) lead to important efficiently gains
through reduction of redundant computations.  Finally, during numerical
phase-space integration, one can introduce importance sampling, computing
computationally expensive but numerically small parts on a subset of phase
space points.

\subsection{Color Decomposition}

At tree-level, amplitudes for $SU(N_c)$ gauge theory with $n$
external gluons can be decomposed into color-ordered partial amplitudes,
multiplied by an associated color trace (see e.g.~\cite{TreeReview}).
Summing over all non-cyclic permutations gives the full amplitude,
\begin{equation} \A{n}^\tree(\{k_i,h_i,a_i\}) = \sum_{\sigma\in S_n/Z_n}
	\Tr(T^{a_{\sigma(1)}} \cdots T^{a_{\sigma(n)}}) \
	A_n^\tree(k_{\sigma(1)}^{h_{\sigma(1)}},\ldots,
	k_{\sigma(n)}^{h_{\sigma(n)}})\ , \label{TreeAmplitudeDecomposition}
\end{equation} 
the coupling is set to one, and $S_n/Z_n$ is the set of non-cyclic permutations of
$\{1,\ldots, n\}$. The $T^a$ are the set of hermitian generators of the
$SU(N_c)$ color group. The coefficients of the color structures $ \Tr(T^{a_{1}}
\cdots T^{a_{n}})$ define the tree-level color-ordered partial amplitudes,
$A_n^\tree(1,2,3,\cdots n)$. 

One of the important features of this set of amplitudes is that it forms a
closed set under collinear, soft and multi-particle factorization. They have
manifest transformation properties under parity transformation and reversal of
the ordering of external legs.  Similarly, amplitudes with fermions can be
decomposed into color-ordered sub-amplitudes~\cite{ColorOrderingFermions}. 

For one-loop amplitudes, one may perform a similar color decomposition to the
one at tree-level in  \eqn{TreeAmplitudeDecomposition}.  In this case, there
are up to two traces over color matrices~\cite{BKColor},
\begin{equation} \A{n}^\oneloop\L \{k_i,h_i,a_i\}\R =
	\sum_{c=1}^{\lfloor{n/2}\rfloor+1} \sum_{\sigma \in S_n/S_{n;c}}
	\Gr_{n;c}\L \sigma \R\,A_{n;c}(\sigma), \label{LoopColorDecomposition}
\end{equation}
where ${\lfloor{x}\rfloor}$ is the largest integer less than or equal to $x$.
The leading color-structure factor $ \Gr_{n;1}(1) = N_c\ \Tr\L T^{a_1}\cdots
T^{a_n}\R $ is just $N_c$ times the tree color factor, and the sub-leading color
structures are given by the double trace expressions, $ \Gr_{n;c}(1) = \Tr\L
T^{a_1}\cdots T^{a_{c-1}}\R\, \Tr\L T^{a_c}\cdots T^{a_n}\R$.  $S_n$ is the set
of all permutations of $n$ objects, and $S_{n;c}$ is the subset leaving
$\Gr_{n;c}$ invariant.

The leading partial amplitudes $A_{n;1}$ take the form (see
e.g.~\cite{OneLoopReview}),
\begin{equation} A_{n;1}^\oneloop(1,2,3,\cdots,n)=A_n^{\rm g}(1,2,3,\cdots
	,n)+\frac{n_f}{N_c} A_n^{\rm f}(1,2,3,\cdots ,n)\,, \end{equation}
with $A_n^{\rm g}$ and $A_n^{\rm f}$ being color-ordered sub-amplitudes, or
primitive amplitudes. While $A_n^{\rm g}$ is fixed to have only gluons
propagating in the loop,  $A_n^{\rm f}$ is restricted to have a Weyl fermion
propagating in the loop. The external gluons are color-ordered; the diagrams
contributing to the loop amplitudes can be generated from color-ordered Feynman
rules, see e.g. ref.~\cite{OneLoopReview}.

The coefficients of the sub-leading color structures; the sub-leading partial
amplitudes, can be expressed in terms of
sums~\cite{Neq4Oneloop,TwoQuarkThreeGluon,BKColor} of the primitive amplitudes,
$A_n^{\rm g}$ where different ordering of the external states appear.
Beyond the fact that such a decomposition exists, we will not need details of
its form here.

\begin{figure}
\begin{center}
\centerline{
\epsfxsize 4 truein\epsfbox{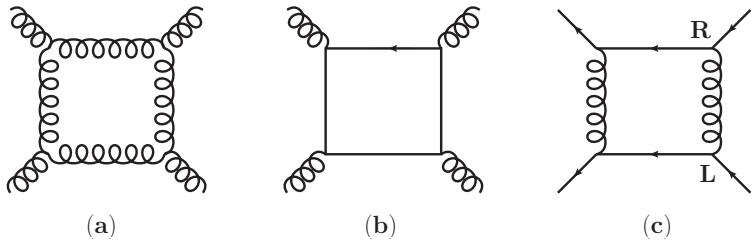}
}
 \caption{
 Representative Feynman diagrams for primitive amplitudes with distinct states
 propagating in the loop: (a) shows a gluon loop of the primitive amplitude
 $A_4^{\rm g}$, (b) a fermion loop  of $A_4^{\rm f}$ and (c) a mixed
 fermion/gluon loop of $A_4^{\rm LR}$.  For the mixed amplitudes (c) we keep
 track of the routing of the fermion line around the loop; '$L\,R$' indicates
 that the first fermion lines turns left and the second right when entering the
 loop.}  \label{PrimitiveAmpls} \end{center} \end{figure}

The primitive amplitudes, 
\begin{equation} 
A_n^{\rm g}(1,2,3,\cdots ,n)\,\quad \mbox{and}\quad\,  A_n^{\rm f}(1,2,3,\cdots ,n)\,, 
\end{equation}
form a generating set of amplitudes, such that given these amplitudes, the full
one-loop matrix elements can be computed.  For fundamental fermions a similar
split-up of partial amplitudes is typically more
complicated~\cite{TwoQuarkThreeGluon,Zqqgg}.  In addition to the ordering of
the external leg the routing around the loop (left- or right-turner) of each of
the fermion lines has to be specified. See figure~\ref{PrimitiveAmpls} for
examples of primitive amplitudes including also external fermion lines.

Primitive amplitudes, have a more transparent analytic structure than full
matrix elements, because their external colored legs follow a fixed cyclic
ordering.  In particular, properties of factorization and branch cut
singularities simplify as can be summarized by the following:
\begin{enumerate}
  \item Only factorization poles and branch cut singularities in adjacent legs appear.
  \item Primitive amplitudes and color-ordered tree amplitudes form a closed
	  set under factorization and unitarity cuts.
\end{enumerate}
With the notation specification of 'adjacent legs' we refer to the fact that
factorization poles and unitarity cuts appear only in a specified subset of
kinematic invariants $s_{i_1\cdots i_k}$, with an ordering of momenta identical
to the one of external gluons.  Closure under factorization means that color
ordered amplitudes factorize onto color ordered amplitudes.  In particular, for
unitarity approaches as well as on-shell recursions, primitive loop amplitudes
can be constructed from color-ordered tree amplitudes alone.

For a more complete description of color decomposition we refer the reader to
 previous reviews~\cite{TwoQuarkThreeGluon,Zqqgg,TreeReview}.

 \subsection{Structure of Loop Amplitude}
\label{StructureOfAmplitude}

\begin{figure}[t]
\centerline{
\epsfxsize 4 truein\epsfbox{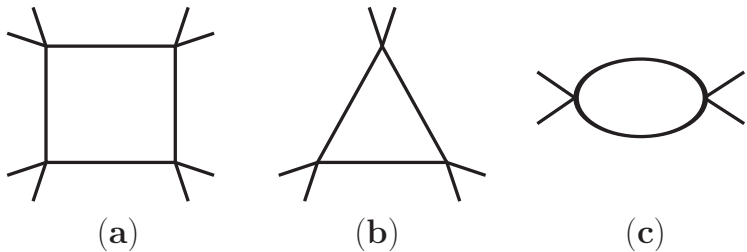}
}
\caption{
Feynman diagram representation of basis of scalar integrals appearing in the
\eqn{IntegralBasis}: 
(a) box diagram associated to the 4-point integral, 
(b) triangle diagram associated to the 3-point integral, and 
(c) bubble diagram associated to the 2-point integral.
Each corner may have one or more external lines attached to it.
} \label{fig:IntegralBasis} \end{figure}

Any one-loop amplitude can be written as a sum of terms containing
branch cuts in kinematic invariants, $C_n$, and a rational remainder $R_n$,
\begin{equation}
A_n = C_n + R_n\,.
\label{CutRational}
\end{equation}
The cut-containing part $C_n$ can in turn be written as a sum over a basis of
scalar master integrals~\cite{IntegralReductions,NVBasis},
\begin{equation} 
C_n =  \sum_{i_1<i_2<i_3<i_4} d^0_{i_1i_2i_3i_4} \,\I_4^{i_1i_2i_3i_4} +
\sum_{i_1<i_2<i_3} c^0_{i_1i_2i_3} \,\I_3^{i_1i_2i_3} + \sum_{i_1<i_2} b^0_{i_1i_2}
\,\I_2^{i_1i_2}\,.
%
\label{IntegralBasis} \end{equation}
The scalar integrals $\I_{2,3,4}$ -- bubbles, triangles, and boxes -- are known
functions~\cite{IntegralsExplicit}.  We omitted tadpole functions, which in
dimensional regularization vanish for massless particles circulating in the
loop.  The explicit dimension dependence is contained in the integral functions
with their coefficient functions strictly four dimensional.  Feynman diagrams
of the integral functions are shown in \fig{fig:IntegralBasis}.

The computation of a one-loop amplitude amounts to determining the rational
coefficient functions $d^0_i,c^0_j$ and $b^0_k$\footnote{As a shorthand we often
specify the collections of indices $\{i_1,i_2,\ldots\}$ by a single one; e.g.
$d^0_i$ instead of $d^0_{i_1i_2i_3i_4}$.} in addition to the remainder $R_n$.  
Following the spinor-helicity method~\cite{SpinorHelicity,Wplus2ME}, we can
then re-express all external momenta in terms of spinors.  The coefficients of
these integrals, $d^0_i, c^0_j$, and $b^0_k$, as well as the remainder
$R_n$, are then all rational functions of appropriate spinor and momentum
variables.

For the analysis of one-loop amplitudes it is often useful to have two distinct
forms of the integrals in mind.  One can think of the integral functions as
logarithms and polylogarithms of kinematic invariants.  As examples we give
explicitly a bubble integral function,
\begin{eqnarray} 
	\I_2(s_{12},\hat\mu^2) &=& { i \, \cg}\biggl\{\frac{1}{\e}+ \biggl(2 +
	\ln\biggl({\hat\mu^2\over-s_{12}}\biggr)\biggr\}+ \Ord(\e)\,,
\end{eqnarray}
exposing discontinuities in kinematic invariants through branch cut
singularities of the underlying logarithmic functions. For the kinematic
invariants $s_{ij}$ the standard definitions, $s_{ij}=(k_i+k_j)^2$, are used.
The constant $\cg$ is defined by,
$ \cg = {\Gamma(1+\eps)\Gamma^2(1-\eps)/( (4\pi)^{2-\eps} \Gamma(1-2\eps))} \,.
$
The scale $\hat\mu$ corresponds to the renormalization and factorization scales
which, for convenience, are set to be equal here; $\hat\mu=\mu_R=\mu_F$. The
integrals are also viewed as Feynman amplitudes of a scalar field theory,
\begin{eqnarray} 
\I_2(s_{12},\hat\mu^2) &=& \hat\mu^{2\e} \int {d^{4-2\e}\ell \over
(2\pi)^{4-2\e}} {1 \over \ell^2 (\ell-K_{12})^2} \,,
\end{eqnarray}
where sums of two momenta are denoted by the shorthand, $K_{ij}=k_i+k_j$.  

\subsection{The Loop Integrand}
\label{TheLoopIntegrand}

For explicit computations it is useful to consider loop amplitudes before
integration, that is to find a universal parametrization ~\cite{OPP,GKM} of
the loop integrand.  In addition to the scalar Feynman diagrams
\fig{fig:IntegralBasis}, implied from \eqn{IntegralBasis}, tensorial numerator
terms have to included.  These tensorial terms describe as well the angular
distribution of the virtual particles, which averages out upon integration.
The explicit relation between numerator tensors and angular variables will be
topic of \sect{AngularDependence}. 

A generic form of the loop integrand is given by~\cite{OPP,GKM} (see
also~\cite{EGK}),
\begin{eqnarray}
A^{\oneloop,d}_n(\ell)&=&
\sum_{i_1<\ldots<i_5}\frac{\bar
e_{i_1i_2i_3i_4i_5}(\ell)}{D_{i_1}D_{i_2}D_{i_3}D_{i_4}D_{i_5}} + \sum_{i_1<
i_2<i_3<i_4}\frac{\bar d_{i_1i_2i_3i_4}(\ell)}{D_{i_1}D_{i_2}D_{i_3}D_{i_4}}
\nonumber\\
&&+\sum_{i_1< i_2<i_3}\frac{\bar c_{i_1i_2i_3}(\ell)}{D_{i_1}D_{i_2}D_{i_3}}
+\sum_{i_1<i_2}\frac{\bar
b_{i_1i_2}(\ell)}{D_{i_1}D_{i_2}}\,,
%
\label{dIntegrandBasis}
\end{eqnarray} 
where $d$ stands for a given discrete dimension and we
restricted external momenta and polarizations to be four-dimensional.  If we
allow $d$-dimensional polarization vectors and external momenta, higher
polygons have to be added in a natural way.  The pentagon terms $\bar e(\ell)$
should be dropped when working in strictly four dimensions, $d= 4$.  In the
above expression, propagators are denoted by $i/D_j=i/(\ell-K_j)^2$; for
simplicity, of notation we restrict the discussion on massless internal states.
Furthermore, we omitted the single propagator terms $\bar
a(\ell)$ which drop out of the final results in the absence of massive states.
When used with the explicit argument $\ell$ for the loop momentum,
$A^{\oneloop,d}_n(\ell)$ denotes the integrand as opposed to the loop amplitude
$A^{\oneloop,d}_n$.

The numerators $\bar e(\ell), \bar d(\ell)$, $\bar c(\ell)$, $\bar b(\ell)$ and
$\bar a(\ell)$ are sums of tensors contracted with the loop momentum
$\ell^\mu$.  The tensor-rank is bounded by power-counting.  We will refer to
these tensors contracted with loop-momentum $\ell$ somewhat imprecisely as {\it
numerator tensors}.  These numerator tensors can be expanded in terms of a
basis of tensors in momentum space multiplied by scalar loop-momentum
independent coefficients.  The scalar coefficients then characterize the loop
amplitude. See below in~\sect{ExplicitTensorBasis} for an explicit
representation in terms of a basis of tensors and scalar coefficients.

Integrand parametrizations (\ref{dIntegrandBasis}) are common in unitarity
approaches; for a discussion in the context of multi-loop computations see
e.g. refs.~\cite{CJReview}.  A particularly useful parametrization of the one-loop
integrand has been given in~\cite{OPP,EGK,GKM}, as will be discussed further
in the following section.


\subsubsection{Loop Integration.}
\label{LoopIntegration}
With an appropriate representation of the loop integrand the loop integrations
can be performed trivially.  This is achieved by writing the integrand
numerators as a direct sum~\cite{OPP} of terms that integrate to zero and non-vanishing
scalar terms.  The form of the integrand in eq.~(\ref{dIntegrandBasis}) then
directly relates to the from~\eqn{IntegralBasis}.  When we evaluate the known
analytic expressions for the basis of integrals, we thus obtain an exact
numerical algorithm to go from an off-shell integrand to the integrated loop
amplitude.  An approach of this kind was used in
refs.~\cite{OPP,BlackHatI,EGK,GKM}. We will motivate a canonical form for the
loop integrand in this section leaving a more complete discussions to the next.
Such a choice can be viewed as an implicit integral reduction procedure.  

To have an example in mind, consider the box numerators in the form,
\begin{eqnarray} 
	\bar d_{i_1i_2i_3i_4}(\ell)&=&d^0_{i_1i_2i_3i_4}+d^{1}_{i_1i_2i_3i_4}
	(n^1\cdot\ell)\,.  \label{BoxNumerator} 
\end{eqnarray}
Here the vector $n^1_\mu$ is understood to be orthogonal to the external
momenta of the box function.  (For an explicit definition and properties of
$n^1_\mu$ see later in \eqn{MomentumBasis}.) The coefficients
$d^0_{i_1i_2i_3i_4}$ and $d^{1}_{i_1i_2i_3i_4}$ are the free parameters of the
ansatz that have to be determined.  The coefficients we eventually need to
compute are the coefficients of the scalar term $d^0_{i_1i_2i_3i_4}$ which
correspond to the scalar basis integral coefficients in~\eqn{IntegralBasis}.
The tensor coefficient $d^{1}_{i_1i_2i_3i_4}$, although necessary at
intermediate steps of the computation drops out after integration,
\begin{eqnarray}
\int \frac{d^4\ell}{(2\pi)^{4-2\ve}}\frac{(n^1\cdot
\ell)}{D_{i_1}D_{i_2}D_{i_3}D_{i_4}}=0\,.
\end{eqnarray} That is, after integration the numerator loop momentum gets
replaces by a linear combination of the external momenta $K_i$ which are
orthogonal to $n^1$; giving a vanishing tensor integral.

It is instructive to consider another form of the tensor numerator, including a
term with an inverse propagator,
\begin{eqnarray} 
	\bar d_{i_1i_2i_3i_4}(\ell)&=&d^0_{i_1i_2i_3i_4}+d^{1}_{i_1i_2i_3i_4}
	(n^1\cdot\ell)+d^{2}_{i_1i_2i_3i_4}D_{i_1}\!(\ell)\,.  
\end{eqnarray} 
Clearly the inverse propagator may be cancelled against one of
the box propagators turning $d^{2}_{i_1i_2i_3i_4}$ into the coefficient of a
scalar triangle function. If we were to treat this term at the box level, we
obtain tensorial contributions that have to be transcribed into scalar integral
functions with some care.  Assuming for the moment some prior knowledge about
generalized unitarity cuts we can make some further observations.
That is, this term actually vanishes on the quadruple cut, leaving the
coefficient $d^{2}_{i_1i_2i_3i_4}$ unspecified at first. It may then be fixed
using the triple-cut equations, although only the sum of the scalar triangle
and $d^{2}_{i_1i_2i_3i_4}$ may be fixed. The numerical unitarity
relations~\eqn{NumericalCuts} are then not triangular. Then box, triangle and
bubble coefficients cannot be solved for consecutively. This is of course
obvious in the present example, given that we rewrote a scalar triangle
coefficient in box-form. This observation emphasises the need for a good basis
of numerator tensors.

A less trivial deformation of the numerator tensors would be to mix in a
propagator term with the linear tensor,
\begin{eqnarray} 
    \bar d_{i_1i_2i_3i_4}(\ell)&=&d^0_{i_1i_2i_3i_4} +d^{1}_{i_1i_2i_3i_4}\biggl(D_{i_1}\!(\ell) + (n^1\cdot\ell)\biggr)\,.  
\end{eqnarray}
Here again a scalar triangle contribution is pulled back into the box integral.
This time the triangular nature of the cut-equations (see below
in~\eqn{NumericalCuts} for the explicit form of the cut-equations) stays intact
and no redundancy is introduced into the numerator tensors. However, one has to
pay attention not to drop the coefficient when integrating the tensor box
integrals. We can read off the box coefficient directly, $d^0_{i_1i_2i_3i_4}$,
and in addition, we have to include $d^{1}_{i_1i_2i_3i_4}$ to the related
triangle coefficient.

Terms may be moved around between integral functions in this way, effectively
introducing a change of basis of integral functions. As a non-trivial
application, a cut completion, that is a subtraction of gram-determinant poles,
may be achievable in this way.  The form of the numerator tensors has important
implication; it allows to keep the unitarity relations triangular, and, keeps
the integration of the loop integrand simple.


\subsubsection{Numerator Tensors.}
\label{SpuriousNumerators}
In the numerical unitarity approach one is naturally lead to obtain equations
for the integrand expression~\eqn{dIntegrandBasis}.  It is then convenient to
use an explicit form of the numerators in terms of a basis of tensors.
Computing a loop amplitude then amounts to determining the free tensor
coefficients.  

There are several natural requirements~\cite{OPP} for a good basis of numerator
tensors.  The first requirement is, that numerator tensors should of course be
general enough to parametrize the loop-integrand we are interested in.
Typically one uses all tensors up to a given rank, as determined by
power-counting.  Furthermore, optimally one would like to use a minimal set of
tensors.  A last requirement is then, that it should be easy to relate the
integrand basis back to the integral representation in ~\eqn{IntegralBasis}.
It turns out that an optimal tensor basis can be found, which satisfies all
the above requirements~\cite{OPP,EGK,GKM}.

For numerator tensors in strictly $d$ dimensions the tensor basis looks
particularly simple. (We will discuss the $D$-dimensional generalizations below
in~\sect{DdimUnitarity}.)  In fact, the result will be a basis of tensors,
called spurious numerators in the literature~\cite{OPP}, which integrate to
zero,
\begin{eqnarray} 
0&=&\hat\mu^{2\varepsilon}\int\frac{d^{4-2\varepsilon}\ell}{(2\pi)^{4-2\varepsilon}}
\frac{n_{\mu_1\cdots\mu_k}\ell^{\mu_1}\cdots \ell^{\mu_k}}{\ell^2
(\ell-K_1)^2\cdots(\ell-K_1\cdots-K_{n-1})^2}\,,\quad k>0\,,
\label{TensorIntegral}
\end{eqnarray}
where $n_{\mu_1\cdots\mu_k}$ stands for a representative basis tensor.  Upon
integration, the loop-momentum dependent numerators in \eqn{dIntegrandBasis}
may thus be dropped and the remaining scalar (rank-zero) terms are directly
identified with the integral coefficients in (\ref{IntegralBasis}).

In order to obtain this basis of tensors it is convenient to introduce the
Neerven-Vermaseren basis~\cite{NVBasis} for vectors in momentum space; a
distinct basis for each of the integral functions.  Each integral defines a
distinguished set of momenta; the momenta $K_i^\mu$ in \eqn{TensorIntegral}.
Momentum space is decomposed into the direct sum of two subspaces; the {\it
physical space} parametrized by $K_i^\mu$ and its complement, spanned by the
vectors $n^{i\mu}$~\cite{EGK},
\begin{eqnarray} 
%
K_i\quad\mbox{for}\quad i\in \{1,\cdots ,n-1\}\,,\quad\quad 
n^i\quad\mbox{for} \quad i\in\{1,\cdots ,d-(n-1)\}\,,\nonumber\\ 
n^i\cdot n^j=\delta^{ij}\,,\quad n^i\cdot K_j=0\,, \nonumber\\ 
v^i\quad\mbox{for}\quad i\in \{1,\cdots ,n-1\}\,, \qquad K_i\cdot
v^j=\delta_i^j\,,\quad n^i\cdot v^j=0\,, \label{MomentumBasis} 
\end{eqnarray}
where we assumed $n\leq d$. For the complementary case $n>d$ momentum space is
parametrized solely in terms of a linearly independent set of vectors
$K_i^\mu$.

The vectors $v^{i\mu}$ are dual to the external momenta $K_i^\mu$ \footnote{ We
may obtain the dual vectors using $\kappa_{ij}=K_i\cdot K_j$ and it's inverse
$(\kappa^{-1})^{ij}$, such that $v^i=\sum_j (\kappa^{-1})^{ij}K_j$. An explicit
form of the dual basis can be found in~\cite{EGK}.  } and are part of the
physical space, defined by the external momenta $K_i^\mu$.  The vectors
$n^{i\mu}$ are an orthonormal basis in {\it transverse space} and are orthogonal to the
physical space. Depending on the signature that the transverse space inherits
from momentum space, the vectors $n^{i\mu}$ have to be chosen purely real or
imaginary.  A further useful quantity is the metric $g_{\perp}^{\mu\nu}$ of the
transverse space,
\begin{eqnarray} 
(g_\perp)^{\mu_1\mu_2}\equiv \sum_{i=1}^{d-n+1}  n^{i\mu_1}n^{i\mu_2}=
g^{\mu\nu} -\sum_{i=1}^{n-1} v^{i\mu} K_i^{\nu}\,, \label{TraceRelation}
\end{eqnarray} 
which is naturally related to the metric of momentum space.

A generic numerator tensors can be expressed as tensor products of the
vectors~(\ref{MomentumBasis}). A basis of tensors is thus given by,
\begin{eqnarray}
n_{\mu_1\cdots\mu_k}=K_{\mu_1}^{i_1}\ldots K_{\mu_l}^{i_l}\,\, n^{j_1}_{\mu_1} \ldots
n^{j_m}_{\mu_m}\,,\qquad l+m=k\,.  \label{GenericTensor}
\end{eqnarray}
In fact, the set of all numerator tensors needed is given by the symmetric
traceless tensors in the transverse space,
\begin{eqnarray} 
&&n_{\mu_1\cdots\mu_k}=n^{i_1}_{\{\mu_1} n^{i_2}_{\mu_2} \cdots
n^{i_k}_{\mu_k\}}\,,
\label{SymTracelessTensors} \end{eqnarray} 
where the curl-brackets denote the operation of symmetrization and subtraction
of traces within transverse space. By trace we mean the contraction,
\begin{eqnarray} 
n_{\mu_1\mu_2\cdots \mu_n}g_\perp^{\mu_1\mu_2}=0\,, 
\end{eqnarray} 
of Lorentz indices with the metric tensor in transverse space,
$g_\perp^{\mu\nu}$.  For symmetric tensors it is sufficient to single out two
indices for contraction. 

Again the case of a vanishing transverse space $n>d$ is special. For this we
are left only with the rank-zero scalar numerators. In fact, for this case a
further simplification appears; the scalar $n$-point integral can be written in
terms of lower-point integrals. To show this one uses identities implied by
inserting vanishing Gram determinants, $\Delta_{d+1}(K_1,\cdots K_d,\ell)=0$,
into $n$-gon integral. Repeated reasoning along these lines leads to reduce
$n$-gon integrals to $d$-gons or lower. We refer to a recent discussion on this
in~\cite{GKKTwoLoop} for further details.

Examples of symmetric traceless tensors are then, $n^{\{1}_\mu
n^{2\}}_\nu=(n^1_\mu n^2_\nu+n^2_\mu n^1_\nu)/2$, which is traceless due to the
orthogonality of the vectors $n^i$. A further example is the tensor,
$n^{\{1}_\mu n^{1\}}_\nu=(n^1_\mu n^1_\nu-(g_{\perp})_{\mu\nu}/(d-n+1))$, for
which the trace was explicitly subtracted.

The form (\ref{SymTracelessTensors}) of the spurious terms can be understood in
the following way. To start with, it turns out that tensors
(\ref{GenericTensor}) with components pointing along the physical space, i.e.
$l>0$, are redundant.  For the simplest case $n^\mu=K_1^\mu$ (with $l=1$) the
contraction of $K_1$ with the loop momentum $\ell$, 
\begin{equation} 
	\ell\cdot K_1=\frac{1}{2}\left[\ell^2-(\ell-K_1)^2+K_1^2\right]\,,
	\label{InversePropagatorRel}
\end{equation}
gives rise to inverse propagators $\ell^2$ and $(\ell^2-K_1)^2$,  and a scalar
term $(K_1^2)$. Although we started with a rank-one tensor integrand, after the
inverse propagators is cancelled, we obtain lower-point integrals and a scalar
integral,
\begin{eqnarray}
\frac{K_1\cdot\ell}{\ell^2 (\ell-K_1)^2 D_2\cdots D_n}&=&
\frac{1}{2}\Biggl[ \frac{1}{(\ell-K_1)^2 D_2\cdots D_n}-
\frac{1}{\ell^2 D_2\cdots D_n}+ \,\nonumber\\
&&+\frac{K_1^2}{\ell^2 (\ell-K_1)^2D_2\cdots D_n}\Biggr]\,.
\label{RedundantTensorIntegral}
\end{eqnarray}
The tensor integrand we started with is thus redundant, as it can be expresses
solely in terms of lower-rank and lower-point terms.  A similar reasoning,
applied recursively, shows the redundancy of tensors with multiple
$K_i$-components and enforces $l=0$ in the notation of \eqn{GenericTensor}.
Tensors including physical directions $K_i$ are thus redundant; they either
lead to linear combination of lower-point integrals or are expressed as tensors
of lower rank. We thus do not need to consider these tensors further, once we
account for this. 

What remains to be considered are tensors with components purely in the
transverse space. Of these only a subset of tensors is linearly independent.
In particular, a trace-containing term in the transverse space is related to a
trace in physical space in addition to a metric tensor.  Thus, a
trace-containing term yields inverse propagators and loop-momentum independent
terms when contracted with loop momentum,
\begin{eqnarray} 
\sum_{i=1}^{d-n+1} n^i_\mu n^i_\nu\,\ell^\mu  \ell^\nu &=& \ell^2 -
\sum_{i=1}^{n-1} (v^i\cdot\ell) (K_{i}\cdot\ell)\,, 
\end{eqnarray}
where we used equation (\ref{TraceRelation}). The contractions of
$(K_{i}\cdot\ell)$ can be transcribed into inverse propagators and terms
independent of loop momentum as in \eqn{InversePropagatorRel}. Traces thus lead
to lower-point or lower-rank tensors and are thus linearly dependent.  The only
choice left are traceless, symmetric tensors in the transverse space.

One might still worry that additional hidden relations can be found to relate
integrals with distinct tensor numerators of the from
(\ref{SymTracelessTensors}). No further relations can in fact be found.  The
independence of the tensor of the same propagator structures can be argued
using an explicit on-shell loop-momentum parametrization in~\eqn{CutMomenta} as
we will discuss further in~\sect{AngularDependence}. The independence of tensor
integrals (\ref{SymTracelessTensors}) with distinct propagator
structures is due to their differing factorization properties; e.g. triangle
integrals with numerator tensors (\ref{SymTracelessTensors}) cannot mimic the
quadruple cut divergence of four-point integrals.


Tensor integrals with symmetric traceless numerator tensors integrate to zero.
Due to Lorentz- and parity-invariance, see e.g. \cite{OPP,CachazoSharpening}, a
generic tensor integral is written in terms of productions of the vectors
$K_i^\mu$ and metric tensors $g_{\mu\nu}$, 
\begin{eqnarray} 
&&\hat\mu^{2\varepsilon}\int\frac{d^{4-2\varepsilon}\ell}{(2\pi)^{4-2\varepsilon}}
\frac{ \ell^{\mu_1}\cdots \ell^{\mu_k}}{\ell^2
(\ell-K_1)^2\cdots(\ell-K_1\cdots-K_{n-1})^2}=\,\nonumber\\ &&\hskip 3.cm=
K_{i_1}^{\{\mu_1}\ldots K_{i_l}^{\mu_l} g^{\mu_{l+1}\mu_{l+2}}\ldots
g^{\mu_{k-1}\mu_{k}\}} f(s_{ij})+\cdots \,. \label{VanishingTensorIntegral}
\end{eqnarray} 
The integrals simply depend on no other vectors and $\epsilon$-tensors are
excluded by parity invariance. Upon contraction with a symmetric traceless
tensor, $n_{\mu_1\cdots \mu_k}$ from (\ref{SymTracelessTensors}), the left hand
side of (\ref{VanishingTensorIntegral}) turns into a tensor integral
(\ref{TensorIntegral}) while the right hand is easily seen to vanishes;
$n_{\mu_1\cdots \mu_k}$ is traceless and the vectors $n^{i\,\mu}$ being
orthogonal to $K_i^\mu$. We thus verified that symmetric traceless numerator
tensors (\ref{SymTracelessTensors}) lead to vanishing integrals
(\ref{TensorIntegral}).

In summary, the symmetric traceless tensors (\ref{SymTracelessTensors}) fulfill
all criteria of an optimal basis as discussed in the beginning of this section.


\subsubsection{Tensor Basis.}
\label{ExplicitTensorBasis}

An explicit form of the numerator tensors in \eqn{dIntegrandBasis} in terms of
the vectors (\ref{MomentumBasis}) was given in~\cite{EGK}. For box
coefficients we have,
\begin{eqnarray}
\bar d_{i_1i_2i_3i_4}(\ell)&=&d^{0}_{i_1i_2i_3i_4}+d^{1}_{i_1i_2i_3i_4} t_1,
\label{NumeratorTensors} 
\end{eqnarray} 
for triangles, 
\begin{eqnarray} 
	\bar c_{i_1i_2i_3}(\ell)&=& c^{0}_{i_1i_2i_3} +
	c^{1}_{i_1i_2i_3}t_1+c^{2}_{i_1i_2i_3}t_2+c^{3}_{i_1i_2i_3}(t_1\,t_1-t_2\,t_2)
	\nonumber\\
	&&+t_1\,t_2\biggl(c^{4}_{i_1i_2i_3}+c^{5}_{i_1i_2i_3}t_1+c^{6}_{i_1i_2i_3}t_2\biggr),
\label{NumeratorTensors3} 
\end{eqnarray} 
and bubbles, 
\begin{eqnarray} \bar
	b_{i_1i_2}(\ell)&=&b^{0}_{i_1i_2}+b^{1}_{i_1i_2}t_1+b^{2}_{i_1i_2}t_2+c^{3}_{i_1i_2}t_3\,\nonumber\\
	&&+b^{4}_{i_1i_2}(t_1\,t_1-t_3\,t_3)+
	b^{5}_{i_1i_2}(t_2t_2-t_3\,t_3)\nonumber\\
	&&+b^{6}_{i_1i_2}t_1\,t_2+b^{7}_{i_1i_2}t_1\,t_3+b^{8}_{i_1i_2}t_2\,t_3\,,
\label{NumeratorTensors2} 
\end{eqnarray}
respectively. Here we introduced $t_i=(n^i\cdot\ell)$. The vectors $n_j$ differ
between the three equations and are defined for each associated propagator
structure individually as defined in~\eqn{MomentumBasis}.  The coefficients we wish
to compute are the $d^{0}_{i_1i_2i_3i_4}$, $c^{0}_{i_1i_2i_3}$ and
$b^{0}_{i_1i_2}$ terms which correspond to the scalar basis integral
coefficients. The tensorial expressions vanish upon integration but have to be
kept at intermediate steps of the computation. 

The above representation is not unique; not only may one chose a different
basis for the transverse space and thus different basis vectors $n^j$. One may
also alter the tensor basis used in this parametrization.  For example, the
above expression uses terms of the form $(t_1\,t_1\,t_2)$ which are not
traceless. An alternative, traceless symmetric representation would be instead 
$(t_1\,t_1\,t_2-t_2\,t_2\,t_2/3)$.  The difference of the two approaches
amounts to moving loop-momentum tensors between bubble- and
triangles-functions.  Given that both forms integrate to zero, either of the
above choices leads directly to the same final result.  (For a related
discussion see also~\sect{LoopIntegration}.)

\subsection{Spinor Helicity}

Spinor variables give a unified way to express polarization vectors of gluons,
fermion helicity states and kinematics of a scattering process.  Furthermore,
spinor variables lead to a natural way to work with complex momenta.
Complex-valued on-shell momenta are important in order to fully exploit
analyticity properties of amplitudes. We will see examples of this in
computations of integral coefficients,
sections~\ref{TripleCutExample} and \ref{QuadrupleCutExample}, and, later
in~\sect{OnShellRecursionSection}, when we consider on-shell recursions.

We follow the standard spinor helicity notation and conventions as
in refs.~\cite{SpinorHelicity,TreeReview}. As a shorthand notation for the
two-component (Weyl) spinors we use,
\begin{equation}
(\lambda_i)_\alpha \equiv [u_+(k_i)]_{\alpha} \,, \qquad
(\tlambda_i)_{\dot\alpha} \equiv [u_-(k_i)]_{\dot\alpha} \,, \qquad k_i^2 =0\,.
\label{lambdadef} \end{equation}
Lorentz-covariant spinor products of left- and right-handed Weyl spinors can be
defined using the antisymmetric tensors $\ve^{\alpha\beta}$ and
$\ve^{\dot\alpha\dot\beta}$, 
\begin{eqnarray} \spa{j}.{l} = \ve^{\alpha\beta} (\lambda_j)_\alpha
	(\lambda_l)_\beta,\qquad \spb{j}.{l} =
	\ve^{\dot\alpha\dot\beta} (\tlambda_j)_{\dot\alpha}
	(\tlambda_l)_{\dot\beta}\,.
	\label{spinorprodtwo} \end{eqnarray}
These products are antisymmetric, $\spa{j}.{l} = - \spa{l}.{j}$, $\spb{j}.{l} =
- \spb{l}.{j}$.  

One can reconstruct the momenta from the spinors,
using $u(k)\bar{u}(k) = \ksl$,
\begin{equation}
	k_i^\mu (\sigma_\mu)_{\alpha\dot\alpha}
	= (\ksl_i)_{\alpha\dot\alpha}
= (\lambda_i)_\alpha (\tlambda_i)_{\dot\alpha} \,.
\label{kfact}
\end{equation}
\Eqn{kfact} shows that a massless momentum vector, written as a bi-spinor, is
simply the product of a left-handed spinor with a right-handed one.  In order
to specify on-shell momenta we will often use the abbreviation,
\begin{eqnarray} 
\ell=\lambda_i (\Ksl\lambda_j)\,, \label{MomentumSpinorRelation}
\end{eqnarray}
where spinorial indices are suppressed and the index contractions are indicated
by the parenthesis;
$(\Ksl\lambda_j)_{\dot\alpha}=\Ksl_{\dot\alpha\alpha}\lambda_j^\alpha$.  Spinor
products of the above momentum $\ell$ are then given by,
$\spa{n}.{\ell}=\spa{n}.i$ and $\spb{\ell}.{n}=\spab{j}.{K}.{n}$.

The usual momentum dot products can be constructed from the
spinor products using the relation,
\begin{equation} \spa{l}.{j} \spb{j}.{l} = 2k_j\cdot k_l = s_{jl} \,.
	\label{spaspbeq} \end{equation}
We will also use the notation,
\begin{eqnarray} 
\spab{a}.{K_{i_1\ldots i_m}}.{b} &=& \sum_{k=1}^{m} \spa{a}.{i_k} \spb{i_k}.{b} \,,\quad\mbox{and}\quad 
s_{i_1\cdots i_m} = K_{i_1\cdots i_m}^2 \, .\label{sjlmdef}
\end{eqnarray}

A further important class of quantities are Gram determinants
$\Delta_n(K_1,\cdots, K_n)$ defined by,
\begin{eqnarray}
	\Delta_n(K_1,\cdots, K_n)\equiv {\rm det}( 2\,K_j\cdot K_l )\,.
	\label{DefGramDet}
\end{eqnarray}
Gram determinants appear naturally in unitarity cuts; when solving for on-shell
momenta negative powers of Gram determinants appear. These then enter the
computation of integral coefficients.  

Gram determinants can be associated to the linearly independent momenta of
integral-functions.  The respective integral coefficients typically have
inverse powers of these Gram determinants in addition to the ones inherited
from reduction of higher-point tensor integrals.  For loop-level on-shell
recursions, \sect{SpuriousResidues}, we will see that Gram determinants play an
important role.

\subsubsection{Basic Tree Amplitudes.}

When using the spinor-helicity formalism tree-level scattering amplitudes
simplify significantly.  Further simplifications arises in part from symmetry
properties of tree
amplitudes~\cite{TreeSymmetries,Integrability,amplitudestructure} that are
present in QCD-like theories (see also~\cite{BSTReview}).  Numerical
implementations of on-shell recursions may recurse all the way to three-point
vertices. More efficiently, they can easily be combined with a library of
compact analytic trees. The recursion is stopped and analytic expressions are
used whenever available, leading to an efficient numerical algorithm.

One example of simplifications due to the use of spinor helicity variables are the
infinite set of vanishing tree-level gluon amplitudes,
\begin{equation} A_n^\tree(1^{\pm},2^{+}, \ldots,n^+) = 0\,,\quad n>3\,,
\end{equation}
with all helicities identical, or all but one identical. Parity may of course
be used to simultaneously reverse all helicities.

The infinite set of Parke-Taylor amplitudes~\cite{ParkeTaylor,Nair} is another
striking example for which the use of spinor helicity formalism yields a
particularly simple form, 
\begin{equation} A^\tree(1^-,2^+,\ldots, j^-,\ldots,n^+)= i\, { {\spa{1}.{j}}^4
	\over \spa1.2\spa2.3\cdots\spa{n}.1 }\ , \quad n\geq3\,, \end{equation}
with two negative helicities and the rest positive. The only gluons with
negative helicity are in positions $1$ and $j$.  Helicities are assigned to
particles with the convention that they are outgoing.
The parity conjugate amplitudes may be obtained by exchanging the left- and
right-handed spinor products in the amplitude, $\spa{j}.{l} \leftrightarrow
\spb{l}.{j}$.

Furthermore, implicit supersymmetry properties~\cite{SusyWardID} allow to
relate fermion, gluon and scalar amplitudes of differing spins.  For example,
in order to replace the gluons $1$ and $n$ by scalar states,
\begin{equation} 
A^\tree(1^-_s,2^+,\ldots, j^-,\ldots,n^+_s)= i\, \frac{\spa{1}.{n}^2}{
{\spa{1}.{j}}^2} {{\spa{1}.{j}}^4  \over \spa1.2\spa2.3\cdots\spa{n}.1 }\ ,
\quad n\geq3\,, 
\end{equation}
we simply multiply the pure gluon amplitude with an overall factor.  Such
relations can be used to speed up the sums (\ref{ColorHelicitySum}) over
final-state particles when using trees with manifest supersymmetry properties
. (See e.g.~\cite{TreeAmplitudes,ManifestSusyTrees} for trees with manifest
supersymmetry properties.)

\subsubsection{On-shell Momenta.}

For real momenta, $\lambda_i$ and $\tlambda_i$ are complex conjugates of each
other up to a sign depending on the sign of the energy component.  For the
degenerate but important case of three-point kinematics,
\begin{eqnarray}
k_!+k_2+k_3=0\,,
\qquad\mbox{with}\quad 
k_i^2=0\,,
\label{OnShellThreePoint}
\end{eqnarray}
only the trivial solution $k_i\sim k_j$ can be found for real momenta. For
these real solutions all spinor products vanish.

However, for complex momenta, it
is possible to choose all three left-handed spinors to be proportional,
$\tlambda_1 = c_1 \tlambda_3$, $\tlambda_2 = c_2 \tlambda_3$, while the
right-handed spinors are not proportional, but obey the relation, $c_1
\lambda_1 + c_2 \lambda_2 + \lambda_3 = 0$, which follows from momentum
conservation, $k_1+k_2+k_3 = 0$. Then, 
\begin{equation} 
	\spa{i}.{j}\neq 0,\quad\mbox{ but}\quad \spb{i}.{j}=0\,.
	\label{threeptcomplex}
\end{equation}
A second branch
of solutions to the on-shell conditions can be found as the conjugate set of
momenta, $\lambda_i\leftrightarrow\tilde\lambda_i$.

Such degenerate kinematics are important for unitarity cuts associated to
integral functions with massless corners.  An explicit computation will be
discussed in \sect{TripleCutExample} and \sect{QuadrupleCutExample}. For such
cases three-point tree amplitudes,
\begin{equation}
A_3^\tree(1^-,2^-,3^+)= i { {\spa1.2}^4 \over \spa1.2 \spa2.3 \spa3.1 } \,,
\label{ThreePointAmpl} \end{equation}
have to be evaluated on solutions to the on-shell conditions
(\ref{OnShellThreePoint}). They are non-trivial on one set of complex solutions
and vanish as $0^4/0^3$ on the other.  The above non-trivial solutions
involving complex momenta are then necessary in order to exploit generalized
unitarity cuts. The general form of this type of on-shell conditions is
discussed below~\eqn{ConeSolutions}.

\subsection{Supersymmetric Decomposition}
\label{SUSYDecomposition}

The supersymmetric decomposition of the amplitudes is particularly useful when
considering rational terms of scattering amplitudes.  In particular, the
rational parts of amplitudes with gluon and fermion degrees of freedom
circulating in the loop can be related to often easier to compute scalar ones.

From power-counting arguments we known a priori that the supersymmetric
amplitudes $A_{n}^{\,\NeqFour}$ and  $A_{n}^{\,\NeqOne \; {\rm chiral}}$ are
cut constructible in four dimensions and are free of rational
terms~\cite{UnitarityMethod}.  The $\NeqFour$ multiplet and the $\NeqOne$
chiral matter multiplet are built from a particular combination of gluon,
fermion and scalar degrees of freedom.  For the case of external gluons, the
couplings of the matter particles resembles the one of QCD, leading to the
relations,
\begin{eqnarray} 
  A_{n}^{\,\NeqFour} &&\;\equiv\; A_{n}^{g}\; +\; 4A_{n}^{f}\;+\;3 A_{n}^{s}\,,\nonumber
  \\ A_{n}^{\,\NeqOne\; {\rm chiral}}&&\;\equiv\; A_{n}^{f}\; +\; A_{n}^{s} \,,  
  \label{susyrelations}
\end{eqnarray}
between supersymmetric amplitudes (lhs) and basic field theory amplitudes (rhs). The
superscripts, $g,f$ and $s$,  indicate the states circulating in the
loop, and stand for gluon, Weyl fermion and a complex scalar, respectively.
Although the above relations are for adjoint fermions in the loop, they can
be directly related to massless fundamental quark
loops~\cite{FiveGluon,UnitarityMethod}.  

Inverting the above relations (\ref{susyrelations}) one obtains the amplitudes for QCD via 
\begin{eqnarray}
  A_{n}^{g} &=& \; A_{n}^{\,\NeqFour}-4A_{n}^{\,\NeqOne\; {\rm
  chiral}}\;+\;A_{n}^{s}\,, \nonumber\\ 
  A_{n}^{f} &=& \; A_{n}^{\,\NeqOne\; {\rm
  chiral}}\;-\;A_{n}^{s}\,.  
  \label{SusyQCDDecomp} 
\end{eqnarray}
This then implies that the rational terms within $A_{n}^{g}$ and
$A_{n}^{f}$ equal the ones from $\pm A_{n}^{s}$, 
\begin{eqnarray} 
  A_{n}^{g}|_{\rm rational}=A_{n}^{s}|_{\rm rational}\,,\quad
  A_{n}^{f}|_{\rm rational}=-A_{n}^{s}|_{\rm rational}\,. 
  \label{ScalarDecomposition}
\end{eqnarray}
With this decomposition we can then compute
the cut containing pieces in strictly four dimensions taking into account the
full QCD spectrum in the loop. At the same time one may compute the rational
part of the QCD amplitudes purely from amplitudes with a complex scalar in the
loop.

When computing the rational terms using the $D$-dimensional unitarity
approach~\cite{DdimUnitarity}, virtual scalars are much more straightforward to
deal with as opposed to gluons or fermions.  While the kinematics of internal
scalars has to be considered beyond four dimensions, we do not have to worry
about $D$-dimensional extension of polarization states of gluons and fermions.
Computations are then very similar to having a massive
scalar~\cite{DdimUnitarity,Badger,DdimUnitarityRecent} in the loop, however,
where the mass is related to the $(D-4)$-dimensional momentum.

Relations of this kind are rather generic and can be found for internal
fermions and mixed fermion and gluon amplitudes. For analytic computations
this observations were used for example in refs.~\cite{TwoQuarkThreeGluon,Zqqgg}.  


\section{The Unitarity Method}
\label{UnitaritySection}

The modern unitarity method~\cite{UnitarityMethod} in addition to generalized
unitarity~\cite{Zqqgg,BCFUnitarity} are the foundation of powerful approaches
for loop computations with phenomenological interest.  Many recent
generalizations~\cite{OPP,EGK,BlackHatI,GKM}, in particular, with a numerical
application in mind, have helped to established a standard unitarity algorithm.
These numerical unitarity methods were first applied to studies of hadron
collider physics in~\cite{PRLW3BH,EMZW3Tev,W3jDistributions,MZ3j,TeVZ}, and are
by now used by many groups~\cite{OtherUnitarityNLO}. Beyond this, various other
implementations of numerical unitarity approaches have been
reported~\cite{GKW,OtherLargeN,SAMURAI,Madloop,BBU}.

Below we will focus on key developments of the unitarity method with
emphasise on numerical aspects. We will follow aspects of the approaches
outlined in~\cite{EGK,BlackHatI,GKM}.  For a discussion of analytic unitarity
methods we refer to the chapter in the review by Britto~\cite{BReview} and
references therein.  A more detailed account of the modern unitarity approach
as well as its application for multi-loop computations may be found in the
chapters of this review by Bern and Huang~\cite{BHReview} and by Carrasco and
Johansson~\cite{CJReview}.

\subsection{Unitarity Relations}

In terms of the non-forward part $T$ of the S-matrix, $S=(1+i\,T)$, unitarity
conditions $SS^\dagger=1$, imply the nonlinear equations,
\begin{eqnarray}
-i(T-T^\dagger)=T^\dagger T.  \label{UnitarityRelation}
\end{eqnarray}
When combined with analyticity properties, as present in field theory, the
unitarity condition (\ref{UnitarityRelation}) relates branch cut
discontinuities of scattering amplitudes, to integrals of products of
scattering amplitudes. (See e.g. refs.~\cite{SMatrixBooks} for an early account of
unitarity and analyticity.) At one-loop order the unitarity relations may be
written as,
\begin{equation} 
{\rm Disc}(s_i)\, A_n^\oneloop=\sum_{\rm states}\int d\Phi(\ell_1,\ell_2)\,
A_{n_1}^\tree(-\ell_2,\ell_1) A_{n_2}^\tree(-\ell_1,\ell_2)\,,
\label{DoubleCut}
\end{equation}
where the state sum is over all intermediate physical states in the theory.
The phase-space integral $\int d\Phi(\ell_1,\ell_2)$, is defined over
integration contours with the intermediate momenta $\ell_1$ and $\ell_2$
on-shell; $\ell_i^2=m_i^2$.  The notation, ${\rm Disc}(s)$, stands for the
branch cut discontinuity in the complexified variable $s$. E.g.  for a
logarithm we have ${\rm Disc}(s)\ln(s/\mu^2)=2\pi i$ such that the operator
${\rm Disc}(s)$ picks out the coefficient of the logarithm.  Importantly, the
nonlinear unitarity relation links on-shell amplitudes of different loop-order
in perturbation theory.  For simplicity, we restrict our discussion to color
ordered amplitudes.

For field-theory amplitudes Cutkosky~\cite{Cutkosky} generalized
\eqn{DoubleCut} further, providing a prescription to directly compute more generic
discontinuities~\cite{Landau}.  An early version of generalized unitarity for generic
field theories was demonstrated in~\cite{Zqqgg,BCFUnitarity}, including
massless states and an arbitrary number of external particles. 
Specialized to one-loop, the discontinuities
are given by phase-space integrals of multiple on-shell scattering amplitudes,
\begin{equation} 
{\rm Disc}(s_{i_1},\cdots s_{i_k})\, A_n^\oneloop=\sum_{\rm states}\ \int
d\Phi\, A^{\tree}_{n_1}A_{n_2}^\tree \cdots {\cal A}_{n_k}^\tree \,.
\label{GeneralizedUnitarityCuts}
\end{equation}
As above, the state sums run over all physical states in the theory. Intermediate
momenta are integrated over appropriate contours of their simultaneous
on-shell phase-space.  

For one-loop computations generalized cuts include single-, double-, triple-,
quadruple- and penta-cuts.  The generalized unitarity cuts are important for
several reasons: first of all, the additional unitarity relations give
further equations to characterize loop amplitudes. Furthermore, the
additional on-shell conditions for the intermediate momenta restrict the
phase-space integral further.  The cuts that are the easiest to evaluate are
the maximal cuts. For these the loop-momenta are frozen to specific values and
the phase-space integral degenerates to only a sum over discrete solutions of the
on-shell conditions.  In four dimensions these correspond to quadruple
cuts~\cite{BCFUnitarity}. Finally, generalized unitarity allows to order cuts
hierarchically; from maximal cuts to next to maximal cuts etc. This hierarchy
allows a systematic approach to inverting unitarity relations. (See below
in~\sect{UnitarityRelationsSection}.)

Applied to Feynman diagrams, the generalized unitarity relations can be formulated
as diagrammatic cutting rules~\cite{Cutkosky}.  In this setup cutting
replaces propagators with on-shell conditions,
\begin{eqnarray} 
\frac{i}{\ell^2-m^2+i\ve}\quad\rightarrow\quad 2\pi\,\delta_{p}(\ell^2-m^2)\,,
\label{CuttingRules}
\end{eqnarray}
and yields directly the values of the associated discontinuities.  (The
subscript 'p' in $\delta_{p}(\ell^2-m^2)$ indicates the common restriction to a
specific branch of on-shell momenta.)
In this way cutting rules (\ref{CuttingRules}) relate unitarity cuts
(\ref{GeneralizedUnitarityCuts}) to universal factorization properties of the
loop integrands. More explicitly, under the loop integral propagators are
replaced by on-shell conditions, such that conditions for the factorization
limits of the loop integrands are obtained,
\begin{eqnarray} 
\lim_{\ell \rightarrow \ell_{i_1\ldots i_k}}\biggl(D_{i_1}\cdots D_{i_k}\,
A^\oneloop_n(\ell)\biggr)=\nonumber\\ \qquad\sum_{\rm
states}\biggl(A^\tree_{n_1}(\ell_{k},\ldots,-\ell_1)\times\cdots\times
A^\tree_{n_k}(\ell_{k-1},\ldots,-\ell_k)\biggr)\,.  
%
\label{NumericalUnitarityRelations} \end{eqnarray}
where $A_n^\oneloop(\ell)$ stands for the loop integrand of the amplitude
$A_n^\oneloop$. The state sums run over the full spectrum of the theory. The
momentum $\ell_{i_1\ldots i_k} $ solves the on-shell conditions $D_{i_1}=\ldots
=D_{i_k}=0$, restricting the loop momentum to the on-shell phase-spaces.  All
intermediate momenta $\ell_i$ are on-shell and related to $\ell$ by momentum
conservation.  The importance of the use of complex momenta for solving the
on-shell conditions for theories with massless states was pointed out in
ref.~\cite{BCFUnitarity} where a algebraic equations for quadruple cuts were
obtained similar to the form of eqn.~(\ref{NumericalUnitarityRelations}). 

\subsection{Inverting The Unitarity Relations}
\label{UnitarityRelationsSection}

We specialize our discussion below on computations in strictly four dimensions.
Analogous ideas generalize to higher dimensions; see \sect{DdimUnitarity}.  In
principle, dispersion integrals~\cite{SMatrixBooks} should allow us to
assemble the amplitude in terms of its branch cut structure, however, for
practical applications a more powerful approach can be
used~\cite{Neq4Oneloop,Zqqgg}.  The unitarity relations may be most directly
implemented by relying on the decomposition of loop amplitudes into a basis of
loop-integral functions (\ref{IntegralBasis}).  Matching the unitarity cuts
with the cuts of basis integrals provides an effective means for obtaining
expressions for the integral coefficients $d_i$, $c_j$ and $b_k$,
\begin{equation}
{\rm Disc}\, {A}_n^\oneloop= \sum_{i} d_i\,{\rm Disc}\,\I^i_4\, +\sum_{j}
c_j\,{\rm Disc}\,\I^j_3\, +\sum_{k} b_k\,{\rm Disc}\,\I^k_2\,.
\label{UnitarityRelations}
\end{equation}
The unitarity cuts of the ansatz are then compared in all channels to the cuts
of the amplitudes.

For a numerical algorithm one aims to further simplify \eqn{UnitarityRelations}
by removing the phase-space integrations. This can be achieved by introducing a
parametrization~\cite{OPP} of the loop integrand (\ref{dIntegrandBasis}). In
addition to scalar integrals, tensor integrals have to be considered.  The
latter parametrize the angular distribution of virtual states circulating in
the loop, as discussed above in~\sect{TheLoopIntegrand}. 

The computation of loop amplitudes then amounts to determining the coefficients
of a basis of scalar and tensor integrands (\ref{NumeratorTensors},
\ref{NumeratorTensors3}, \ref{NumeratorTensors2}) and subsequently performing
the loop integrations.  Coefficients in the ansatz of (\ref{dIntegrandBasis})
may be determined through comparison to factorization limits of loop amplitudes
(\ref{NumericalUnitarityRelations}).  The unitarity method applied at the
integrand level~\cite{Darren,EGK} leads to a set of equations for on-shell loop
momenta,
\begin{eqnarray} 
\bar d_{i_1i_2i_3i_4}(\ell)&=&\sum_{\rm
states}A_{n_{i_1}}^\tree(\ell)A_{n_{i_2}}^\tree(\ell) A_{n_{i_3}}^\tree(\ell)
A_{n_{i_4}}^\tree(\ell)\,,
    \nonumber\\
	\bar c_{i_1i_2i_3}(\ell)&=&\sum_{\rm
	states}A_{n_{i_1}}^\tree(\ell)A_{n_{i_2}}^\tree(\ell)
	A_{n_{i_3}}^\tree(\ell)-\sum_{j\neq i_1,i_2,i_3}\frac{\bar
	d_{i_1i_2i_3j}(\ell)}{D_j} \,,
    \nonumber\\
    \bar b_{i_1i_2}(\ell)&=&\sum_{\rm
    states}A_{n_{i_1}}^\tree(\ell)A_{n_{i_2}}^\tree(\ell)-\sum_{j\neq
    i_1,i_2}\frac{\bar c_{i_1i_2j}(\ell)}{D_j}-\sum_{\stack{j,k\neq
    i_1,i_2}{j<k}}\frac{\bar d_{i_1i_2jk}(\ell)}{D_jD_k} \,,
    \label{NumericalCuts} \end{eqnarray} 
where the individual equations have to be evaluated on respective on-shell
momenta; $\ell\rightarrow \ell_{i_1i_2i_3i_4i_5}$ or $\ell\rightarrow
\ell_{i_1i_2i_3i_4}$, etc.  The momentum dependence of the individual tree
amplitudes is indicated with the appropriate shift of the momentum dependence
implicitly assumed.  

Typically, the on-shell conditions do not constrain the loop momenta
completely, but lead to a variety of solutions. Enforcing the unitarity
relations (\ref{NumericalCuts}) on the variety of on-shell solutions gives an
infinite set of equations.  The set of tensorial structures in addition to the
scalar integral coefficients can be determined due to this
degeneracy~\cite{OPP,Darren,EGK,BlackHatI}.

\subsection{Angular Dependence of Numerator Tensors.}
\label{AngularDependence}

It turns out, that the unitarity relations (\ref{NumericalCuts}) allow us to
obtain the coefficients of all scalar and tensor numerator terms, and, thus,
allow to reconstruct the off-shell loop integrand.  This is despite the fact
that the unitarity relations are defined only for on-shell intermediate momenta.
We will discuss a key aspect of this off-shell power of the on-shell unitarity
equations in this section.

An important observation is, that the on-shell conditions from cutting
propagators fix the values of the loop momenta in the physical space, but
introduce almost no restrictions in the transverse space~\cite{EGK}.  Following
the notation in \eqn{MomentumBasis}, an explicit form of the cut loop momentum
is given by~\cite{EGK}, 
\begin{eqnarray}
%
\ell^\mu&=&V^\mu+\sum_{i=1}^{d-n+1} \alpha^i n^\mu_i\,,\quad \sum_{i=1}^{d-n+1}
\alpha^i\alpha^i=-V^2\,, \label{CutMomenta}
\end{eqnarray}
where $V^\mu$ represents a specific linear combination of vectors $\{K_i\}$ and
lies within physical space. The coordinates $\alpha_i$ may take complex values
and parametrize the transverse part of the loop momentum; only the square of
the transverse part is fixed.  We will denote the parameter space,
$\vec{\alpha}\equiv\{\alpha^i\}_{i=1,d-n+1}$ with $\vec{\alpha}^2=-V^2$, by
$M_{\vec{\alpha}}$ in the following.  It turns out that the remaining freedom
in the transverse momentum directions allows us to uniquely specify all scalar
and tensor coefficients in~eqs.(\ref{NumeratorTensors}),
(\ref{NumeratorTensors3}) and (\ref{NumeratorTensors2}). 

The numerator tensors contracted with the on-shell loop momenta
(\ref{CutMomenta}) give linearly independent functions on the parameter space
$M_{\vec{\alpha}}$.  This can be seen considering the definition of the basis
of numerator tensors. For a non-vanishing value of $V^2\neq 0$ the parameter
space contains a $(d-n)$-dimensional sphere, $S_{d-n}\subset M_{\vec{\alpha}}$.
The basis tensors (\ref{SymTracelessTensors}) are symmetric and traceless in
the transverse vectors $n^{i\,\mu}$. Thus, when we contracted a numerator
tensor with the loop-momentum (\ref{CutMomenta}) we use $n^i\cdot
K_j=0=n^i\cdot V$ and obtain harmonic polynomials in terms of $\alpha_i$. These
polynomials are of course linearly independent being in one-to-one
correspondence to spherical harmonics on the respective sphere, $S_{d-n}$. 

For example, for two-particle cuts in four dimensions, $n=2$ and $d=4$, we
may use spherical coordinates for the coordinate vector $\vec{\alpha}$ in
(\ref{CutMomenta}) and obtain the classical spherical harmonics $Y_{lm}$, 
\begin{eqnarray}
&&\vec{\alpha}=i\,|V|\times\{ \cos\theta, \sin\theta\cos\phi,
\sin\theta\sin\phi \}\,,\\ 
&&n^{i_1}_{\{\mu_1}\ldots n^{i_k}_{\mu_k\}}\,\ell^{\mu_1}\ldots
\ell^{\mu_k}\sim Y_{l,m}(\theta,\phi),\quad l=k\,,m=-l,\ldots ,l\,, \label{CutConditions}
\end{eqnarray}
as observed in~\cite{BlackHatI} using a specific representation of the on-shell
momenta. Definitions of curly-brackets and vectors in (\ref{CutConditions}) are
given in eqs.~(\ref{MomentumBasis}) and (\ref{SymTracelessTensors}),
respectively.  The norm $|V|$ is given by $|V|\equiv\sqrt{V\cdot V}$.
Similarly, one obtains a representation in terms of spherical harmonics for
generic dimensions and tensor-rank of the numerator tensors.  

Quadruple cuts in four dimensions, $n=d=4$, are a special case.  For this case
the transverse space is one-dimensional, $\alpha_1=\pm i |V|$, with the
harmonics degenerating to two functions; the even and the odd functions on two
points, $\alpha_1=\pm i |V|$, representing a zero-dimensional sphere, $S_0$.
The even and the odd functions, $Y$ and $Y'$, on these two points can be defined
to take the respective values, $Y=\{1,1\}$ and $Y'=\{1,-1\}$. The constant
function $Y$ corresponds to the scalar integral and $Y'$ to the linear tensor
in ~\eqn{NumeratorTensors}. An analogous behaviour appears as well away from
four dimensions, $d\neq4$, whenever we have $d=n$.

A refinement of the parametrization in terms of spherical coordinates
(\ref{CutConditions}) is necessary when the square of $V^\mu$ vanishes;
$|V|=0$. This case arises for cuts of integrals with massless internal and at
least one massless external leg. The parameter space then takes the form of a
cone,
\begin{eqnarray}
	\sum_{i=1}^{d-n+1}\alpha^i\alpha^i=0\,. \label{ConeSolutions}
\end{eqnarray}
A generic parametrization of (\ref{ConeSolutions}) is given by
$\vec{\alpha}=\{t, i\,t \vec{\beta}\}$ with $t$ and $\vec{\beta}$ complex
valued and $\vec{\beta}^2=1$.  Also in this case the numerator tensors have a
unique functional form in terms of $t$ and spherical harmonics originating from
angular dependence of $\vec{\beta}$.  In particular, they are linearly
independent functions on the parameter space $M_{\vec{\alpha}}$.

A special case worth mentioning are triple cuts with massless external legs;
$d=4$ and $n=3$.  For this situation we have only a two dimensional transverse
space, $\alpha_i$ with $i=1,2$, and the angular dependence through
$\vec{\beta}$ degenerates further to $\beta_1=\pm1$.  The solution space
to~\eqn{ConeSolutions} then consists of two branches, $\vec{\alpha}=\{t, \pm
i\,t\}$.  A simple example of this situation is given in
\sect{TripleCutExample}. Both branches of phase-space have to be taken into
account to solve the unitarity relations.  

In summary, we see that the tensor numerators in \eqn{dIntegrandBasis} are 
in one-to-one correspondence with a basic set of functions (e.g. spherical
harmonics) that appear in the unitarity cut. In particular, all tensor
coefficients may be identified using the linear independence of these functions.
To this end, we may evaluate the unitarity relations (\ref{NumericalCuts}) on a
predefined set of points on the parameter spaces $M_{\vec{\alpha}}$ and use the
known functional form of each of the numerator tensors to pick out their
coefficients.  The degrees of freedom left after the on-shell conditions are
imposed, are thus sufficient to uniquely obtain the loop integrand through
unitarity cuts.

\subsubsection{Power Counting.} 
A further interesting observation is, that the
total angular momentum quantum number that appears, i.e. $l$ of $Y_l$, is
constrained by the power-counting of the theory. For terms with $P$ powers of
loop momentum $\ell^\mu$ we have,
\begin{eqnarray}
Y_{l}\,,\quad l\leq P \quad\mbox{for maximal power of}\quad (\ell^\mu)^P\,.  
\end{eqnarray}
Conversely, the maximal angular momentum reflects the UV behaviour of the
respective amplitude. For the special case of $d=n$ this reasoning does not
apply.  For simplicity, we focused only on the situation with $|V|\neq 0$ and
we only assert that a similar counting argument can be extended to the case
$|V|=0$.

The total number of numerator tensors can then be counted using representation
theory of the orthogonal group, $SO(d-n+1)$, and are given by the number of
harmonics up to a given total angular momentum.

\subsection{A Triple-Cut Example}
\label{TripleCutExample}

\begin{figure}[t]
\centerline{
\epsfxsize 4 truein\epsfbox{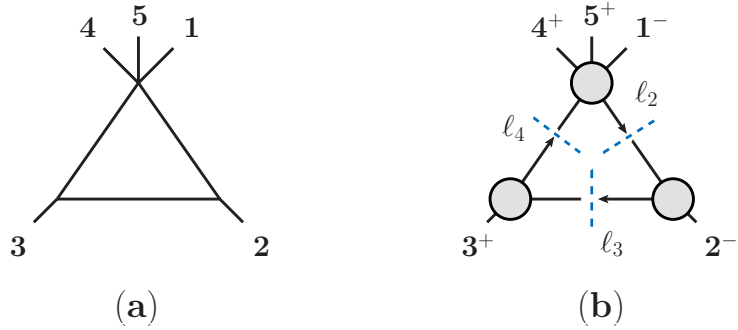}
}
\caption{The triangle integral function $\I_3(k_2,k_3,K_{451})$ (left) and the associated
triple cut (right). The virtual states in the loop are fixed to be a complex scalar.
While the scalar integral coefficient turns out to vanish, some tensor
coefficients are non-zero. The tensorial coefficients are important input for
the computation of lower-point integral coefficients in the numerical unitarity
approach.} \label{TripleCutFigure} \end{figure}

We consider the amplitude $A^{s}_5(1^-,2^-,3^+,4^+,5^+)$ with a scalar
circulating in the loop.  Amplitudes of this kind were first computed in
refs.~\cite{UnitarityMethod} using the unitarity method.  Here we are only
interested in the computation of a single triple-cut, as shown
in~\fig{TripleCutFigure}.  It turns out that some tensor coefficients
(\ref{NumeratorTensors}) are non-vanishing while the scalar triangle
coefficient $c_{234}^0$ vanishes. This simple example allows to illustrate how
to apply some of the methods that were introduced earlier.

In later parts of this review we will discuss further important aspects of such
unitarity computations.  In \sect{DiscreteFourier}  we will discuss, based on
the present example, how to compute tensor coefficients numerically using
discrete Fourier transformation, following the notation in~\cite{BlackHatI}.
Building on this, in \sect{NumericalStability}, we will explain how numerical
precision can be monitored during computation of such triple-cuts.

We start with the computation of the triple cut~\fig{TripleCutFigure}. A
momentum parametrisation and reference momenta, similar
to~\eqn{MomentumBasis}, are given by,
\begin{eqnarray} 
	&&n^+=\lambda_2\tilde\lambda_3\,,\quad
	n^-=\lambda_3\tilde\lambda_2\,,\quad n^\pm\cdot K_i=0\,,\nonumber\\
&&\ell_2^a=\left(\lambda_2+t\lambda_3\right)\,\tilde\lambda_2\,,
\quad \ell_3^a=t\,\lambda_3\,\tilde\lambda_2\,,\quad
\ell_4^a=\lambda_3\,\left(-\tilde\lambda_3+t\tilde\lambda_2\right)\,.
\label{TripleCutMomenta}
\end{eqnarray} 
This is a special case of solutions of the on-shell conditions
for integrals with massless corners as explained below \eqn{CutConditions}.
The on-shell momentum space takes the form of a light cone and consists of two
parts, with the second branch, $\ell_i^b$, related to (\ref{TripleCutMomenta})
by parity reflection $\lambda^i\leftrightarrow \tilde\lambda^i$. The loop
momentum is then,
\begin{eqnarray} 
\ell_2^b=\lambda_2(\tlambda_2+t'\tlambda_3)\,\quad
\ell_3^b=t'\,\lambda_2\tlambda_3\,,\quad
\ell_4^b=(-\tlambda_3+t'\tlambda_2)\tlambda_3\,.  
\end{eqnarray} 
Notice that a
more standard way to write the above parametrization would use $n^{1}=i(n^++
n^-)/\sqrt{s_{23}}$, $n^{2}=(n^+- n^-)/\sqrt{s_{23}}$ and $\ell_3^{a,b}=
-t\sqrt{s_{23}}(i n^1\pm n^2)$.  The use of complex momenta is crucial in order
to evaluate such unitarity cuts.

The triple-cut, \fig{TripleCutFigure}, is given by the product of three tree amplitudes,
\begin{eqnarray} 
\hskip-1cm \bar c_{234}(\ell_i)&=&
	A_3^\tree((-\ell_2)^s,2^-,(\ell_3)^s) 
	A_3^\tree( (-\ell_3)^{s},3^+,\ell_4^{s})
	A_5^\tree((-\ell_4)^{s},4^+,5^+,1^-,\ell_2^{s})\,. 
\end{eqnarray}
Inserting explicit tree amplitudes we then obtain,
\begin{eqnarray}
	\bar c_{234}(\ell_i)=
	\frac{(-i)\spb3.{\ell_4}^2\spb3.{(-\ell_3)}^2  }{\spb{3}.{\ell_4}\spb{\ell_4}.{(-\ell_3)}\spb{(-\ell_3)}.{3}}\,\times\nonumber\\
	\frac{i\, {\spa{(-\ell_4)}.1}^2{\spa{\ell_2}.1}^2}{\spa{(-l_4)}.4\spa4.5\spa5.1\spa1.{\ell_2}\spa{\ell_2}.{(-l_4)}}\times
	\frac{ i\,{\spa{(-\ell_2)}.2}^2 {\spa{\ell_3}.2}^2}{\spa{(-\ell_2)}.2\spa2.{\ell_3}\spa{\ell_3}.{(-\ell_2)}}\,,
\end{eqnarray}
and after using the above momentum parametrisation, \eqn{TripleCutMomenta},
\begin{eqnarray}
\bar c_{234}(\ell_i^a)&=&-\frac{t^2\spa1.3^2\spb3.2\left(\spa2.1+t\,\spa3.1\right)}{\spa3.4\spa4.5\spa5.1}\,,
\nonumber\\
\bar c_{234}(\ell_i^b)=0\,,
\label{TripleCutEval}
\end{eqnarray}
where the triple-cut vanishes on the second branch $\ell^b$ due to the vanishing
of the three-point vertices for these momenta.

We can make several observations: (a) There is no term constant in $t$
consistent with the absence of scalar triangles~\cite{UnitarityMethod}. (b) The
maximal power of $t$ is three, consistent with the three powers of loop
momentum expected from power-counting of a triangle diagram. (c) The
expression (\ref{TripleCutEval}) has no poles at finite non-zero values of $t$
consistent with the absence of scalar boxes for this particular combination of
external helicities and scalar states in the loop~\cite{UnitarityMethod}. The
equations (\ref{NumericalCuts}) thus simplify with box-subtraction terms being
zero and, thus, absent.

The parametrization of the numerator tensors (\ref{ExplicitTensorBasis}) is given by,
\begin{eqnarray}
\bar c_{234}(\ell)&=&
c^{3}\,(n^+\cdot \ell)^3+c^{2}\,(n^+\cdot \ell)^2+c^1\, (n^+\cdot \ell)+c^0+ \,\nonumber\\
&& 
c^{-1}(n^-\cdot \ell)+ c^{-2}(n^-\cdot \ell)^2+c^{-3}(n^-\cdot \ell)^3\,.
\label{TripleNumeratorParametrization}
\end{eqnarray}
The next step is to relate this parametrization of the numerator tensors
(\ref{TripleNumeratorParametrization}) to the expressions for the triple cut
\eqn{TripleCutEval}. For on-shell momenta we find for
(\ref{TripleNumeratorParametrization}),
\begin{eqnarray}
\bar c_{234}(\ell^a)&=&c^{3}\,(-s_{23} t)^3+c^{2}\,(-s_{23}t)^2+c^1\, (-s_{23}
t)+c^0\,,\nonumber\\ 
\bar c_{234}(\ell^b)&=&c^0+c^{-1}\,(-s_{23}
t')^1+c^{-2}\,(-s_{23}t')^2+c^{-3}\,(-s_{23} t')^3\,,
\label{TripleCutAnsatz}
\end{eqnarray}
where any of the on-shell momenta $\ell^{a,b}_i$ $i=2,3$ or $4$, can be used.

Comparing the polynomials in $t$ and $t'$ given by cut (\ref{TripleCutEval})
and ansatz (\ref{TripleCutAnsatz}) we find all tensor coefficients,
\begin{eqnarray}
c^0=0=c^1=c^{-i}\,,\quad c^{i\,\geq\, 2}=-s_{23}\,A_5^\tree
\left(\frac{\spa3.1}{\spa2.1}\right)^i\,.
\end{eqnarray} 
In an analytic approach, it is of course no problem to compare the
$t$-dependent expressions. What matters here, is that the tensor basis gives
linearly independent functions on the solution space of $\ell^{a,b}$, so that
all tensor coefficients can be identified. This will turn out sufficient for
the related numerical approach, as we will discuss in~\sect{DiscreteFourier}.

Thus, effectively by replacing
$t\rightarrow-(n^+\cdot\ell_2)/s_{23}$ in (\ref{TripleCutEval}) we obtain the
off-shell loop-momentum dependence,
\begin{eqnarray}
	\bar c_{234}(\ell)=-s_{23}\, A^\tree_5\,\left( \frac{(n^+\cdot
	\ell)\spa3.1}{\spa2.1}\right)^2 \left(1+
	\frac{(n^+\cdot\ell)\spa3.1}{\spa2.1}\right)\,,
\end{eqnarray}

For the special case of this example, many coefficients
including the scalar coefficient $c^0$ vanish. Because of this the associated
triangle integral does not contribute to the given amplitude.  The computation
of the tensor coefficients is important, nevertheless, as they are used when
computing the two-particle cuts, for which they serve as subtraction terms as
apparent from~\eqn{NumericalCuts}.

In the above analysis we relied on the fact that we had analytic expressions
available. In the following section we will see that a purely numerical approach
can be set up to follow very similar computational steps to obtain scalar and
tensor coefficients.

\subsection{Discrete Fourier Transform} 
\label{DiscreteFourier}
%
\begin{figure}[t]
\centerline{
\epsfxsize 2.5 truein\epsfbox{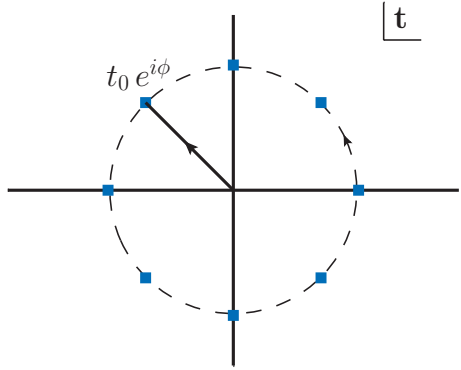}
}
\caption{
The discrete Fourier transformation to obtain integral coefficients. The triple
cut contribution depends only on a small number of functions.  Instead of a
contour integral, the triple-cut may thus be evaluated on a discrete set of
points.  All scalar and tensor coefficients are obtained exactly in a numerical
procedure.  
} \label{TripleCutFourier} \end{figure}
%
We continue the discussion of the above example in \sect{TripleCutExample}. A
direct way to extract the tensor coefficients from the cut expression
$\bar c_{234}(\ell_i^a)$ in \eqn{TripleCutEval} is through contour integrals,
\begin{eqnarray}
c^k=\frac{1}{2\pi i}\int \frac{dt}{t} \left(\frac{t}{-s_{23}}\right)^{-k} \bar
c_{234}(t)\,, 
\end{eqnarray} 
which take the form of a Fourier transformation in an angle $\phi$ for an
integration contour $t=t_0 e^{i\phi}$.

From a power-counting argument we know a priory that only a finite number of
monomials in $t$ may appear. In particular, we know that $t$ will at most appear as
a third power $t^3$.  We may thus use a discrete version of the above contour
integral~\fig{TripleCutFourier},
\begin{eqnarray}
c^k=\frac{1}{m}\sum_{j=1}^m \left(\frac{t_0 e^{2\pi i\,
j/m}}{-s_{23}}\right)^{-k} \bar c_{234}(t= t_0 e^{2\pi i\, j/m} )\,,  \quad\mbox{with}\quad m\geq4\,,
\end{eqnarray}
giving an exact numerical way to obtain the required tensor coefficients.
Typically the expected tensor-rank determines the number of points that need to
be sampled in the above sum. For the example in \sect{TripleCutExample} four
points, $m=4$ would be sufficient.

For the computation of all tensor coefficients in
equations~(\ref{NumeratorTensors}), (\ref{NumeratorTensors3}) and
(\ref{NumeratorTensors2}) as well as in equations~(\ref{DPentagon}),
(\ref{DBox}), (\ref{DTriangle}) and (\ref{DBubble}) analogous discrete sums
can be constructed~\cite{BlackHatI}. The central observation is that the
numerator tensors are related to a finite number of functions. For phase-space
integrals the orthogonality of these functions can be used to extract a
particular tensor coefficient from a unitarity cut. For numerical approaches a
related  discrete versions of such integrals can be setup.

\subsection{Numerical Stability}
\label{NumericalStability}

A clear understanding of the relation between numerator tensors and a function
basis in phase-space can be further exploited; we can check the numerical
precision during the computation of loop amplitudes. 

Loss of precision may occur on a given phase-space point, for example, due to
an unfortunate choice of reference momenta $n^i$ (\ref{MomentumBasis}) or a
vanishing Gram determinant (\ref{DefGramDet}). A particularly universal
way to deal with loss of precision is the use of higher precision
arithmetics~\cite{CutTools}.  As demonstrated in ref.~\cite{BlackHatI}, for an
efficient approach one tries to identify the computational steps that lead to
large round-off errors and then to re-evaluates only the problematic contributions to
the amplitude (and only those terms) using higher-precision arithmetic.  Use of
higher precision arithmetics is more time consuming.  Such an approach requires
that results be sufficiently stable in the first place, so that the use of
higher precision is infrequent enough to incur only a modest increase in the
overall evaluation time; this is indeed the case.

The simplest test of numerical
stability~\cite{CutTools,EGK,BlackHatI,GZ,Madloop} is checking whether the
known infrared singularity of a given matrix element has been reproduced
correctly. Typically, a combination of various checks is required.  A refined
test~\cite{BlackHatI} can be setup to check the accuracy of the vanishing of
certain higher-rank tensor coefficients.  From power-counting arguments we know
on general grounds which high-rank tensor coefficients have to vanish. 
All tensors with rank greater than $m$ must vanish, for the $m$-point integrals
with $m=2,3,4$.  The central advantage of this check is that small parts of the
computation may be singled out for re-computation, reducing the computation
time accordingly.

For the above example of a triple cut computation, \sect{TripleCutExample}, a
check of the vanishing tensor coefficients may be implemented as an extension
of the discrete Fourier sum. If a tensor of rank-four were to exist it could be
computed numerically by,
\begin{eqnarray}
c^{k=4}=\frac{1}{m}\sum_{j=1}^m \left(\frac{t_0 e^{2\pi i\,
j/m}}{-s_{23}}\right)^{-k} \bar c_{234}(t=t_0 e^{2\pi i\, j/m})\,,
\quad\mbox{with}\quad m\geq 5. 
\end{eqnarray}
Clearly $c^4$ should always turn out to be zero, $c^4=0$. For finite precision
the deviation of $c^4$ from zero can be used to monitor precision of the
intermediate computational steps. This test requires to sample more values of
the complex parameter $t$; $m=5$ instead of $m=4$. However, observing an
instability at the level of an integral coefficient as opposed to a failing
check of an infrared-pole has the big advantage, that one can recompute very
targeted a small part (the specific triangle coefficient) of the full
amplitude. 

The parameters of such a rescue system have to be tuned in order to optimize
efficiency. This includes the question which integral coefficients to
check and what deviations from zero to expect for a given tensor coefficient.
Some further details of this approach can be found, for example,
in~\cite{BlackHatI}.

\subsection{A Box Example.}
\label{QuadrupleCutExample}

\begin{figure}[t]
\centerline{
\epsfxsize 4 truein\epsfbox{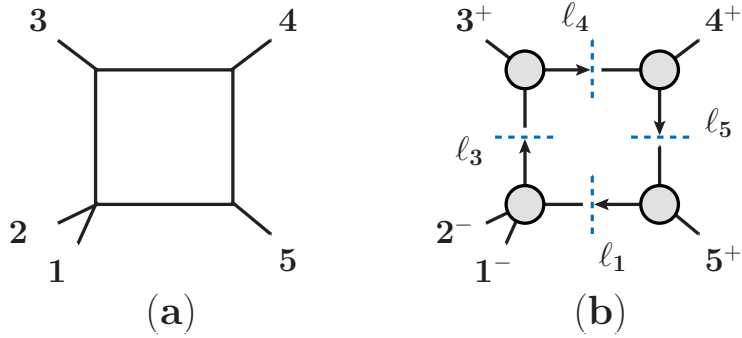}
}
\caption{The box integral function $\I(K_{12})$ (left) and the associated
quadruple cut (right). The helicity sum over internal states is suppressed in
the graphic.} \label{QuadCutFigure} \end{figure}

Here we will discuss the pure gluon amplitude
$A_5^g(1^-,2^-,3^+,4^+,5^+)$.  The analysis below will be purely
analytic, however, in the way we set it up it is straightforward perform all computations numerically.
Since this amplitude has five external massless legs, the box functions
can have one external massive leg, and three massless ones.  One of these box
functions with massive legs $K_{12}=k_1+k_2$, is given by,
$$ 
{\cal I}(K_{12})=\hat\mu^{2\varepsilon}\int
\frac{d^{4-2\varepsilon}\ell}{(2\pi)^{4-2\varepsilon}} \frac{1}{\ell^2
(\ell-k_1-k_2)^2(\ell-k_1-k_2-k_3)^2(\ell+k_5)^2)}\,, 
$$ and displayed in~\fig{QuadCutFigure}.
The other boxes have the massive leg being $K_{23}$, $K_{34}$, $K_{45}$ or
$K_{51}$.  

We focus here on the analysis of the coefficient of ${\cal I}(K_{12})$. The
first step is to parametrize the integrand $\bar d_{1345}(\ell)$ in
\eqn{dIntegrandBasis} that extends the box function with a numerator tensor.
The transverse momentum space (\ref{MomentumBasis}) is one-dimensional and is
spanned by a single vector $n^{1\,\mu}$, 
\begin{eqnarray}
n^1\cdot K_{12}=n^1\cdot k_3=n^1\cdot k_4=n^1\cdot k_5=0\,,
\end{eqnarray} 
with the explicit solution for $n^{1\,\mu}$ given by,
\begin{eqnarray}
n^1\cdot
\ell\equiv([3|k_4k_5\ell|3\rangle-\langle3|k_4k_5\ell|3])/s_{34}s_{45}\,,
\label{BoxBasisVectors}
\end{eqnarray} 
where the loop momentum $\ell$ is identified with $\ell_1$ as defined in
\fig{QuadCutFigure}.
The only symmetric traceless tensor (\ref{SymTracelessTensors}) in the
transverse space is given by the vector $n^{1\,\mu}$ itself; $n^\mu=n^{1\,\mu}$
as in \sect{ExplicitTensorBasis}.  The numerator tensors of the box integrals
are then parametrized as, 
\begin{eqnarray}
&&\frac{d^0_{1345} + d^1_{1345}\,(n^1\cdot\ell)}{\ell^2
(\ell-k_1-k_2)^2(\ell-k_1-k_2-k_3)^2(\ell+k_5)^2)}\,,  \label{BoxAnsatz}
\end{eqnarray}
resembling the generic form in \eqn{NumeratorTensors}. 

The coefficients $d^1_{1345}$ and $d^0_{1345}$ have to be determined by
comparing the quadruple cut equation (\ref{NumericalCuts}) to the ansatz
(\ref{BoxAnsatz}) at the associated on-shell kinematics. In four dimensions the
on-shell conditions,
\begin{eqnarray}
\ell^2&=0=(\ell-k_5)^2=(\ell+k_3)^2=(\ell+K_{34})^2\,,
\label{BoxOnshellMomenta}
\end{eqnarray} 
have two solutions.  A general form for the on-shell momenta can be
found in ref.~\cite{BlackHatI}.  Explicitly, one set of on-shell momenta is given
by,
\begin{eqnarray} 
	\mbox{solution (a):}\quad
	&&\ell_5=\frac{s_{23}}{\spba4.K_{12}.5}\lambda_5 \tlambda_4\,,\quad
	\ell_1=-\frac{\spa3.4}{\spba4.K_{12}.5} \lambda_5
	(\Ksl_{12}\lambda_3)\,,\nonumber\\ &&\ell_3=\frac{\spb3.4}{\spba4.K_{12}.5}
	\lambda_3 (\Ksl_{12}\lambda_5)\,,\quad
	\ell_4=\frac{\spba3.K_{12}.5}{\spba4.K_{12}.5} \lambda_3\tlambda_4\,,
	\label{QuadCutMomenta}
\end{eqnarray} 
with the second set of on-shell momenta, solution (b), being related by parity
conjugation; exchanging $\lambda_i\leftrightarrow \tilde\lambda_i$. We used
here notation as discussed in \eqn{MomentumSpinorRelation}.

The quadruple cut is given by a product of
four tree amplitudes summed over internal helicity states,
\begin{eqnarray} 
\hskip-1cm \bar d_{1345}(\ell_i)&=& \sum_{h_i}A_4^\tree(
	(-\ell_2)^{-h_2},1^-,2^-,\ell_3^{h_3}) A_3^\tree( (-\ell_3)^{-h_3},
	3^+,\ell_4^{h_4}) \times\nonumber\\ &&\qquad\times A_3^\tree(
	(-\ell_4)^{-h_4},4^+,\ell_1^{h_1}) A_3^\tree(
	(-\ell_1)^{-h_1},5^+,\ell_2^{h_2})\,.  
\label{QuadCutExample}
\end{eqnarray}
Only the momenta in \eqn{QuadCutMomenta} give a non-vanishing contribution to
the quadruple-cut (\ref{QuadCutExample}); for the internal helicities
$\{h_1,h_2,h_3,h_4\}=\{-,-,+,+\}$.
In total we have,
\begin{eqnarray} 
	%
\hskip-1cm \mbox{solution (a):}\quad \bar d_{1345}(\ell_i)  &=&
\frac{\spa1.2^3}{\spa2.{\ell_3}\spa{\ell_3}.{(-\ell_2)}\spa{(-\ell_2)}.1}
\frac{\spb3.{\ell_4}^3}{\spb{\ell_4}.{(-\ell_3)}\spb{(-\ell_3)}.3}\times\nonumber\\
&&\quad\times
\frac{\spa{\ell_1}.{(-\ell_4)}^3}{\spa4.{\ell_1}\spa{(-\ell_4)}.4}
\frac{\spb{(-\ell_1)}.5^3}{\spb5.{\ell_2}\spb{\ell_2}.{(-\ell_1)}}
\nonumber\\ &=&
i s_{34} s_{45} \, A_5^\tree(1^-,2^-,3^+,4^+,5^+)\,,\nonumber\\
\hskip-1cm \mbox{solution (b):}\quad \bar d_{1345}(\ell_i)  &=& 0\,.
\label{K12coefficient}
\end{eqnarray}

In order to determine the integral coefficients $d_{1345},d^1_{1345}$ we have to
evaluate ansatz for the integral on the unitarity cut. For the particular
momenta no other terms in the ansatz contribute leading terms in the
factorization limit. We thus obtain, 
\begin{eqnarray} 
	\bar d_{1345}(\ell_i^{a,b})=d^0_{1345} + d^1_{1345}\,(n^1\cdot\ell_i^{a,b})  &=&
	d^0_{1345} \pm d^1_{1345}\,, 
	\label{LinearEquations}
\end{eqnarray} 
where any on-shell momentum $\ell_i$ may be used in the numerator tensor, due
to the properties of the vector $n^{1\,\mu}$ as specified in \eqn{BoxBasisVectors}. 

Using the values of the quadruple cut in \eqn{K12coefficient} we thus can solve
for the unknowns $d^0_{1345}$ and $d^1_{1345}$, 
\begin{eqnarray}
	d^0_{1345}=d^1_{1345}=\frac{i}{2} s_{34} s_{45} \,
	A_5^\tree(1^-,2^-,3^+,4^+,5^+)\,.	\label{QuadCutFinal}
\end{eqnarray}
That the integral coefficients $d^0_{1345}$ and $d^1_{1345}$ turn
out to take identical values is a low point accident related to the presence of
three-point amplitudes in the unitarity cut.  For the value of the one-loop
amplitude the tensorial term is of no immediate importance; it drops out
after integration. However, for the computation of the remaining triangle and
bubble coefficients this is an important ingredient.

The normalization factor $1/2$ in scalar integral
coefficient (\ref{QuadCutFinal}) is usually attributed to an 'averaging over'
quadruple cut solutions~\cite{BCFUnitarity}.  In the presented computation this
factor appears, somewhat differently, from inverting the set of linear
equations (\ref{LinearEquations}).

\subsection{Rational Terms From $D$-dimensional Unitarity Cuts}
\label{DdimUnitarity}

The rational terms (\ref{CutRational}) are free of branch cut singularities and
cannot be detected by unitarity methods in four dimensions.  Away from four
dimensions, $D=4-2\ve$, rational terms carry factors of
$(-s)^{-\e}=(\hat\mu)^{2\e}(1-\e\ln(-s/\hat\mu^2)+\Ord(\e^2))$,
in order to compensate for the dimensionality of the coupling constant. The so
introduced logarithms, in turn, make it possible to detect rational terms via
$D$-dimensional unitarity methods~\cite{DdimUnitarity} (see
also the early work~\cite{vanNeervenDdim}).  This version of unitarity, in which tree
amplitudes are evaluated in $D$ dimensions, has been used in various
analytic~\cite{DdimUnitarityRecent,BFMRational} and
numerical~\cite{GKM,OPPrat,Badger,ggttg,CutTools,GZ,W3EGKMZ,OtherLargeN,HPP}
studies.  For a detailed discussion of analytic $D$-dimensional approaches we
refer to~\cite{BReview}.  Here we will follow the approach discussed
in~\cite{GKM} including elements of~\cite{Badger}.

\subsubsection{Dimension Dependence.}
The prescription of the $D$-dimensional unitarity method is to compute the
$D$-dimensional loop-amplitudes and in the end take the limit, $D=4-2\ve$ with
$\ve\rightarrow 0$. The combined dependence of integrals and their coefficients
on the regulator $\ve$ yields the finite rational terms. 

In numerical approaches the dependence on the dimension parameter $D$ is not
as accessible as in an analytic approaches. Amplitudes may only be computed in
a fixed dimension. The typical strategy, to deal with this issue, is to
numerically compute in distinct discrete dimensions.  With a clear
understanding of the dimension dependence~\cite{GKM}, the dimension parameter
may be reinstated in the final steps of the computation. The limit
$D\rightarrow 4$ is then performed analytically. We will discuss the
$D$-dependence in detail in the following.

The origin of the $D$-dependence is twofold.  While the external momenta and
states are kept in four dimensions, polarization states and momenta in the loop
are extended beyond four dimensions.  The two sources for dependence on $D$ are
the sums over virtual polarization states in $D$ dimensions and momentum
invariants formed using $D$-dimensional loop momentum.  

We consider first the dependence of rational terms on $D$-dimensional
polarization states.  A priori one would need to sum over the particle spectra
of virtual gluons and fermions extended beyond four dimension. This can be done
in practice~\cite{GKM}, however, we will describe a shortcut, that avoids
considering polarizations states beyond four dimensions altogether.  Applying
the arguments of \sect{SUSYDecomposition} we can relate the rational terms of
QCD amplitudes with internal gluons and fermions to amplitudes with virtual
scalar and fermionic states. The extension of scalar states to $D$-dimensions
is straight forward as, in contrast to gluonic states, no additional
polarizations have to be taken into account. In fact, it turns out that the $D$
dependence originating in the change of the number of polarization states with
dimensionality my thus be avoided.  Such an approach is computationally more
efficient.  The computation time grows faster than linearly with the number of
particle states circulating in the loop. The computation of amplitudes of a
complex scalar as opposed to (massless) vector particles leads to efficiency
gains in $D>4$ dimensions.

For the computation of rational terms it is, thus, sufficient to consider
simplified, yet generic amplitudes with: 
\begin{enumerate}
\item closed scalar loops, 
\item mixed scalar and fermion loops.
\end{enumerate} 
Even though QCD has no scalars, introducing them is a useful trick.  For QCD
like theories these assumptions are no restriction for the computation of
rational terms; contributions from either only virtual gluons or fermions are
related to virtual scalars (\ref{ScalarDecomposition}).  Similarly, rational
terms from mixed gluon and fermion loops can be mapped to the ones with the
gluon replaced by a scalar, see e.g.~\cite{TwoQuarkThreeGluon,Zqqgg}.  Loops
with mixed fermion and scalar states do only give rise to state sums over
fermions which exit to external sources.  Since for these amplitudes no closed
loop of Dirac gamma matrices can be formed, no explicit dependence on the
number of fermion states appears.  We thus avoid the dependence on the
dimensionality through the number of spin states.

The second source of the dimension dependence of loop amplitudes is
the dependence on the $D$-dimensional loop momentum. This dependence is
simplified due to the fact that we keep external polarizations and momenta in
strictly four dimensions.  Splitting the loop momentum into a four- and
$(D-4)$-dimensional part, $\ell^\mu =\hat\ell^\mu+\tilde\ell^\mu$, the
dependence on the $(D-4)$-dimensional part is limited to the form,
\begin{eqnarray}
\mu^2=-(\tilde\ell\cdot\tilde\ell)\,.  \label{DMomentumDependence}
\end{eqnarray}
That is, no other vectors with non-vanishing components in $(D-4)$ are
available to form scalar products and we ignore linear $\mu$-terms built from
$\ve$-tensors.  Rotation symmetry in the $(D-4)$-dimensional part of momentum
space is preserved in this way. 

\subsubsection{The Loop Integrand.}
The integrand basis has to include pentagon terms, when
working in $D$-dimensions~\cite{GKM}, 
\begin{eqnarray}
A^{\oneloop,D}_n(\ell)&=&
\sum_{i_1<\ldots<i_5}\frac{\tilde
e_{i_1i_2i_3i_4i_5}(\ell)}{D_{i_1}D_{i_2}D_{i_3}D_{i_4}D_{i_5}} + \sum_{i_1<
i_2<i_3<i_4}\frac{\tilde d_{i_1i_2i_3i_4}(\ell)}{D_{i_1}D_{i_2}D_{i_3}D_{i_4}}
\nonumber\\
&&+\sum_{i_1< i_2<i_3}\frac{\tilde c_{i_1i_2i_3}(\ell)}{D_{i_1}D_{i_2}D_{i_3}}
+\sum_{i_1<i_2}\frac{\tilde
b_{i_1i_2}(\ell)}{D_{i_1}D_{i_2}}\,,
\label{DIntegrandBasis}
\end{eqnarray}
with propagators and numerators depending on the $D$-dimensional loop momentum
$\ell^\mu=\hat\ell^\mu+\tilde\ell^\mu$.  We suppressed the tadpole terms
$a_{i_1}(\ell)$ which are not needed for computation of rational terms in
amplitudes with only massless states. As compared to the earlier expression
(\ref{dIntegrandBasis}) the form of the numerator terms has to be adapted to
the $D$-dimensional case as will be discussed in~\sect{DNumeratorTensors}.  The
generalization to massive states has been presented in~\cite{ggttg}.

The absence of terms with more than five propagators in \eqn{DIntegrandBasis}
is due to the restriction to strictly four dimensional external momenta and
polarization vectors~\cite{NVBasis,IntegralsExplicit}. (See
also~\cite{GKKTwoLoop} for a recent discussion including generalizations to
two-loop integrals.)  Integrands with more than five propagators can be reduced
to at least pentagon terms.  At most five independent momentum vectors can be
formed in $D$ dimensions, since the $(D-4)$-dimensional momentum
$\tilde\ell^\mu$ is conserved in the loop.  Because of this, Gram determinants
which depend on the loop momentum and five or more independent momenta have to
vanish. The resulting identities between momentum invariants can be used to
reduce higher $n$-point propagator structures recursively to at least pentagon
integrals.

\subsubsection{$D$-dimensional Unitarity Relations.}
As for the cut-containing parts, the computation of $D$-dimensional loop
amplitudes amounts to determining the coefficients of a basis of scalar
and tensor integrands and subsequently performing the loop integrations. A
generalization of the unitarity relations (\ref{NumericalCuts}) can be used to
completely determine the loop integrand (\ref{DIntegrandBasis}).

The restriction to purely four-dimensional external momenta has important
implication for the unitarity cuts of the internal propagators.  In fact, the
$(D-4)$-dimensional component $\tilde\ell^\mu$ of the loop momentum is
conserved and enters all propagators in the form of a mass term,
\begin{eqnarray}
\frac{i}{(\ell-K_i)^2}=\frac{i}{(\hat\ell-K_i)^2-\mu^2}\,.
\end{eqnarray}
For unitarity cuts, this observation leads to an important restriction. In
addition to the four momentum components of $\hat\ell^\mu$ we thus obtain only
one additional degree of freedom, i.e.  $\mu^2$.  We can then put at most five
propagators on shell, allowing us to consider at most penta cuts. Of course
this meshes well with the absence of higher point functions
in~\eqn{DIntegrandBasis} in the first place.

The unitarity relations are generalized to $D$ dimensions, by including
an additional generalized penta cut,
\begin{eqnarray} 
\bar e_{i_1i_2i_3i_4i_5}(\ell)&=&\sum_{ D\hbox{-}{\rm dim}\,\,
{\rm states}}A_{n_{i_1}}^\tree(\ell)A_{n_{i_2}}^\tree(\ell) A_{n_{i_3}}^\tree(\ell)
A_{n_{i_4}}^\tree(\ell) A_{n_{i_5}}^\tree(\ell)\,,
    \label{DNumericalCuts} 
\end{eqnarray} 
which is defined for appropriate on-shell loop momentum, $\ell$. The remaining
unitarity cuts are implemented similar to the four-dimensional case
(see~\eqn{NumericalCuts}), however, including additional subtraction terms
from the pentagon-coefficients $\tilde e_{i_1i_2i_3i_4i_5}(\ell)$. 

We may work in any discrete dimension bigger than four, which accounts for the
full loop-momentum dependence in $D$ dimensions.  The internal state sums have
to be extended to the $D$ dimensional setup.  For the case of a scalar state
circulating in the loop, the $D$-dimensional extension of the scalar tree
amplitudes is to be considered.  Given the simple dependence
(\ref{DMomentumDependence}) on the $(D-4)$-dimensional loop momentum it is
sufficient to consider a scalar in five dimensions (see also~\cite{GKM}) to
obtain the full dependence on $\mu^2$.  Equivalently, one can consider
four-dimensional virtual scalar states with a (dynamical)
mass~\cite{DdimUnitarity,Badger}.  

Typically, the on-shell conditions do not constrain the loop momenta
completely, but lead to a variety of solutions. Enforcing the unitarity
relations (\ref{DNumericalCuts}) on the variety of on-shell solutions gives an
infinite set of equations.  We will discuss the parametrization of the
$D$-dimensional numerator tensors below in~\sect{DNumeratorTensors}
and~\sect{DExplicitTensorBasis} which allows to determine all needed integral
coefficients. 

\subsubsection{Numerator Tensors.}
\label{DNumeratorTensors}
In order to parametrize the numerator tensors of a given propagator structure
in~\eqn{DIntegrandBasis} it is convenient to introduce an adapted vector basis 
in momentum space.  The $D$-dimensional momentum space is split into
three subspaces spanned by the vectors $K_i$, $n^i$ and $m^i$~\cite{GKM},
\begin{eqnarray} 
K_i\quad\mbox{for}\quad i\in \{1,\cdots ,n-1\}\,,\nonumber\\ 
n^i\quad\mbox{for} \quad i\in\{1,\cdots ,4-(n-1)\}\, \qquad m^i\quad\mbox{for}
\quad i\in\{5,\cdots ,D\}\,\\
(n^i,n^j)=\delta^{ij}\,,\quad (m^i,m^j)=\delta^{ij}\,,\quad (n^i,m^j)=0\,,
\quad (n^i,K_j)=0\,, \nonumber\\ 
g^{\mu\nu}_{\perp}\equiv\sum_{i=1}^{4-(n+1)}n^{i\mu} n^{i\nu}\,,\quad
g^{\mu\nu}_{D-4}\equiv\sum_{i=5}^{D}m^{i\mu} m^{i\nu}\,. 
\label{DMomentumBasis} 
\end{eqnarray}
In a first step, the $D$-dimensional space is split into two subspaces: the
$4$-dimensional subspace and its $(D-4)$-dimensional complement, ${m^i}$. In a
second step, the $4$-dimensional momentum space is further decomposed into a
physical space, spanned by the independent external momenta of the integral
$K_i$ and their transverse space within four dimensions, $\{n^i\}$. For
convenience the bases $\{n^i\}$ and $\{m^i\}$ are orthonormal. The metrics
$g^{\mu\nu}_{\perp}$ and $g^{\mu\nu}_{D-4}$ are projections of $g_{\mu\nu}$ to
the respective subspaces.

We can follow similar steps as in \sect{SpuriousNumerators} to conclude
that the generic numerator tensors are traceless symmetric tensors in the whole of the
transverse space,
\begin{eqnarray} 
&&n_{\mu_1\cdots\mu_k}= n^{i_1}_{\{\mu_1} \cdots n^{i_l}_{\mu_l}
m^{j_1}_{\mu_{l+1}} \cdots m^{j_k}_{\mu_k\}} \,,\nonumber\\
&&n_{\mu_1\mu_2\cdots \mu_k}g_\perp^{\mu_1\mu_2}+n_{\mu_1\mu_2\cdots \mu_k}g_{D-4}^{\mu_1\mu_2}=0\,,
\end{eqnarray} 
where the curly-brackets indicate symmetrization and subtraction of all traces.
More explicitly, the traces should be subtracted in transverse space, using the
metrics $g_\perp^{\mu\nu}$ and $g_{D-4}^{\mu\nu}$, but not $g^{\mu\nu}$.  Due
to the rotation invariance in $(D-4)$ dimensions, the dependence on the $m^i$
is further restricted to appear solely in terms of
$g^{\mu\nu}_{D-4}=\sum_{i=5}^{D}m^{i\mu} m^{i\nu}$.  

A subtlety appears here for numerical applications; the traceless-condition
introduced a dependence on the dimensionality of space time, as
illustrated by the following example,
\begin{eqnarray}
n_{\mu\nu}=\sum_{i=5}^{D}m^i_{\{\mu}
m^i_{\nu\}}=\biggl( (g_{D-4})_{\mu\nu}-\frac{D-4}{4-(n+1)}\,(g_\perp)_{\mu\nu}\biggr)\,,
\label{TraceExample}
\end{eqnarray}
yielding a $D$-dependent coefficient of $g_\perp^{\mu\nu}$ and thus an explicit
$D$-dependence.  For numerical computations we would like to avoid specifying $D$
and keep it as a parameter.  Without loss of generality, the dependence on
the space-time dimensionality can be tracked more conveniently by allowing terms
with non-vanishing traces.  This can be achieved at the minor cost of
introducing terms that do not integrate to zero, \sect{DIntegration}. 
In particular, as given in~\cite{GKM} we use instead a representation of the
numerator tensors in terms of tensors that are trace-free only in the physical
transverse space, 
\begin{eqnarray} 
&&n_{\mu_1\cdots\mu_k}= n^{i_1}_{\{\mu_1} \cdots n^{i_l}_{\mu_l\}}
g^{D-4}_{\mu_{l+1}\mu_{l+2}} \cdots g^{D-4}_{\mu_{k-1}\mu_{k}}\,,\qquad
n_{\mu_1\mu_2\cdots \mu_k}g_\perp^{\mu_1\mu_2}=0\,,
\end{eqnarray} 
The above rank-two tensor (\ref{TraceExample}) would then be given by,
\begin{eqnarray}
n_{\mu\nu}\ell^\mu\ell^\nu=(g_{D-4})_{\mu\nu}\ell^\mu\ell^\nu=-\mu^2\,,
\end{eqnarray}
without explicit dependence on the parameter $D$.

\subsubsection{Tensor Basis.}
\label{DExplicitTensorBasis}

An explicit form of the numerator tensors was given in~\cite{GKM}.  The
pentagon numerator tensors are,
\begin{eqnarray} \tilde{e}_{i_1i_2i_3i_4i_5}(\ell)&=&e^{0}_{i_1i_2i_3i_4i_5}\,,
	\label{DPentagon} \end{eqnarray}
where the absence of any loop-momentum dependence has to be noted. For this
case the transverse space coincides with the $(D-4)$ dimensional space and no
traceless tensor can be formed. Therefore terms of the form $\mu^2$ can be
converted into inverse propagators and scalar terms and are represented by
lower-point integrals and the scalar pentagon.\footnote{Ignoring $\ve$-tensors
$\mu$-dependent terms of the pentagon can be converted to
propagators and scalars;
$-\mu^2=(g_{D-4})_{\mu\nu}\ell^\mu\ell^\nu=\ell^2-\sum_{i=1,4}
(K_i\cdot\ell)(v^i\cdot\ell)$. Terms of the form $(K_i\cdot\ell)$ can be
further expressed in terms of inverse propagators as in
\eqn{InversePropagatorRel}.}

The remaining numerator tensors are, for the box,
\begin{eqnarray}
\tilde{d}_{i_1i_2i_3i_4}(\ell)&=&\bar d_{i_1i_2i_3i_4}(\ell)+\mu^2(d^{2}_{i_1i_2i_3i_4}+d^{3}_{i_1i_2i_3i_4}t_1)+\mu^4\,d^{4}_{i_1i_2i_3i_4},
\label{DBox}
\end{eqnarray}
for the triangles,
\begin{eqnarray}
\tilde{c}_{i_1i_2i_3}(\ell)&=&\bar c_{i_1i_2i_3}(\ell)+\mu^2\left(c^{7}_{i_1i_2i_3}t_1+c^{8}_{i_1i_2i_3}t_2+c^{9}_{i_1i_2i_3}\right),
\label{DTriangle}
\end{eqnarray}
and finally for bubble coefficients they are,
\begin{eqnarray}
 \tilde{b}_{i_1i_2}(\ell)&=&\bar b_{i_1i_2}(\ell)+\mu^2 b^{9}_{i_1i_2}.
\label{DBubble}
\end{eqnarray}
Here the vectors we introduced $t_i=(n^i\cdot\ell)$. The $n^i$ are defined for
each propagator structure of (\ref{DIntegrandBasis}) individually as defined in
\eqn{DMomentumBasis}.  The tadpole contributions may be found in the original
literature. The $\mu$-independent tensors take the same form as given earlier,
being traceless symmetric tensors in the vectors $n^i$. These tensorial
structures parametrize the most generic tensor integrals, subject only to
power-counting requirements of QCD like theories as well as the constraints
from rotational symmetry manifest in dimensional regularization.

\subsubsection{Integration.}
\label{DIntegration}
The final step to extract the $D$-dimensional amplitude is to evaluate the
integrals. We are interested here only in the computation of the rational
terms.  Only the new integrand structures of eqs.~(\ref{DPentagon}), (\ref{DBox}),
(\ref{DTriangle}) and (\ref{DBubble}), proportional to powers of $\mu$ may
contribute to the rational remainder in eq.~(\ref{IntegralBasis}). 

As above terms dependent on vectors $n^i$ integrate to zero due to their
angular dependence and can be dropped. In order to obtain the rational
contributions we are left with terms proportional only to powers of $\mu^2$. Without
spelling out the details, the non-vanishing limits are given
by~\cite{Mahlon,DdimUnitarity,Badger},
\begin{eqnarray}
\lim_{D\rightarrow 4}\quad   \int
\frac{d^D\ell}{i\pi^{D/2}}\frac{\mu^4}{D_{i_1}D_{i_2}D_{i_3}D_{i_4}}&=&-\frac{1}{6}\,, \quad 
\lim_{D\rightarrow 4}\quad\int
\frac{d^D\ell}{i \pi^{D/2}}\frac{\mu^2}{D_{i_1}D_{i_2}D_{i_3}}=-\frac{1}{2}\,, \nonumber\\
\lim_{D\rightarrow 4}\quad\int
\frac{d^D\ell}{i\pi^{D/2}}\frac{\mu^2}{D_{i_1}D_{i_2}}&=&-\frac{1}{6}\,K_{i_1}^2, 
\label{RationalIntegrals} \end{eqnarray}
The limits of these integrals combined with the associated integral
coefficients add up to the rational term of the one-loop amplitudes.  The
rational term is then given by~\cite{BFMRational,OPPrat,GKM}, 
\begin{eqnarray}
R_n&=&
-\sum_{i_1< i_2<i_3<i_4}\frac{d_{i_1i_2i_3i_4}^4}{6}
-\sum_{i_1< i_2<i_3}\frac{c_{i_1i_2i_3}^9}{2}
-\sum_{i_1<i_2}\frac{K_{i_1}^2\,\, b_{i_1i_2}^9}{6}\,. 
%
\end{eqnarray}

We note that the $D$-dimensional approach is very general and can be applied to
higher loop computations.  For evaluation of scheme shift and order $\ve$
contributions we refer to~\cite{GKM}.


\section{On-shell Recursion}
\label{OnShellRecursionSection}
On-shell recursions~\cite{BCFW} rely on on-shell scattering amplitudes with a
fixed number of partons in order to obtain the ones with arbitrary
multiplicity.  The underlying structures used in this approach are universal
factorization and analyticity properties.  

In fact, when intermediate states are nearly on-shell, amplitudes factorize
into products of lower-point amplitudes.  At tree-level, the naive attempt to
invert factorization equations and assemble the on-shell factorized amplitudes
into their parent amplitude raises two central questions: Firstly, how to
recover the full off-shell kinematics away from the factorization limit? And,
secondly, how to combine various factorization channels while avoiding
double-counting?  On-shell recursion was introduced by Britto, Cachazo, Feng
and Witten~\cite{BCFW}.  These naive obstructions are overcome by the effective
use of complex kinematics and the use of Cauchy's residue theorem.

While originally constructed for tree amplitudes on-shell recursions can be
extended to loop-level~\cite{Bootstrap,Genhel,CoeffRecursion}.  Compared to
tree amplitudes, the obvious difficulty is that loop amplitudes contain branch
cuts, which complicate the use of Cauchy's theorem.  The resolution to these
difficulties has been to work on subparts of the amplitudes, which are purely
rational functions. The price to pay is that additional, un-physical
singularities have to be dealt with, which are present in subparts but cancel
once the full amplitude is assembled.  Un-physical poles can typically be
understood from complementary information.  In the case of on-shell recursions
for rational terms spurious singularities are characterized using prior
knowledge of the logarithmic parts of the amplitude.  In addition, while
factorization of loop amplitudes is understood for real kinematics, presently
there are no theorems on the factorization properties of loop amplitudes with
complex momenta.  Indeed, there is a class of poles, 'unreal poles'
~\cite{Bootstrap}, whose contributions have to be taken into account and whose
nature is not yet fully understood. For most applications unreal poles can be
avoided using alternative factorization channels. (See,
however, ref.~\cite{UnrealPole} for recent progress with understanding the
origin of unreal poles.) 

Most recent developments focus on recursion relations at the integrand
level~\cite{NimaAllLoop} and have already led to many results in maximally
supersymmetric Yang-Mills theory. Here we are mainly interested in on-shell
recursions for loop amplitudes, which can be applied after integration has been
performed.

For numerical applications at tree-level, on-shell recursions are efficient,
allowing to repeatedly take advantage of the remarkable simplicity of the
physical scattering amplitudes (e.g. Parke-Taylor amplitudes).  At loop-level,
on-shell recursions hold the potential of an efficient complement to unitarity
approaches for two reasons: (a) Recursions can sidestep the computation of
tensors coefficients (\ref{dIntegrandBasis}) and deal entirely with the much
smaller set of scalar integral coefficients. (b) On-shell recursions for
rational terms rely on strictly four dimensional information and, thus, do not
require more involved $D$-dimensional objects.  Despite the potential of loop
recursions, unitarity approaches are very universal and straightforward to
implement for a large class of theories explaining their widespread use.
The \BlackHat{}-library~\cite{BlackHatI}, for example, makes use of both
approaches and like this improves the efficiency of computations.

We will first point out some central ideas of on-shell recursions at tree-level
and will then turn to loop-level recursion. 

\subsection{Tree Recursion}
\label{TreeRecursion}

For a more complete discussion on tree level recursions we refer the reader to
the chapter of the present review dedicated to this subject by Brandhuber, Spence
and Travaglini~\cite{BSTReview}. Here we will merely introduce basic notation
needed for the discussion of loop-level on-shell recursions.

On-shell recursion relations systematically construct a scattering amplitude
(or parts of it) from its poles and residues in momentum space.  Typically, a
parametrization of the complexified phase-space in terms of a complex parameter
$z$ is introduced.  An amplitude $A_n(z)=A_n(\{k_i(z)\})$ is then reconstructed
as the unique function with consistent poles and residues in the $z$-plane.
Analyticity is in many cases so restrictive for the form of the amplitudes,
that this is in fact possible.  Prior knowledge of singularities in phase-space
as well as a simple physical interpretation is a necessary input in this
procedure.  

A particularly useful parametrization of a plane in complexified phase-space is
given in terms of a deformation of two momentum vectors~\cite{BCFW},
\begin{equation} A_n(z)\ =\
	A_n(k_1,\ldots,k_j(z),k_{j+1},\ldots,k_l(z),\ldots,k_n).
\end{equation}
The momenta $k_i(z)$ are chosen to keep momenta on-shell and the overall
momentum conservation intact.  The explicit form of such a parametrization,
denoted by $\Shift{j}{l}$, is given by the linear transformation,
\begin{equation}
 \Shift{j}{l}:\quad \tlambda_j \rightarrow \tlambda_j - z\tlambda_l \,,
 \quad\quad \lambda_l \rightarrow \lambda_l + z\lambda_j \,,
 \label{SpinorShift}
\end{equation}
where $z$ is a complex parameter.  This shift of spinors leaves untouched
$\lambda_j$, $\tlambda_l$, and the spinors for all the other particles in the
process.  The corresponding momenta are,
\begin{eqnarray}
k_j(z) = k_j - z \,\lambda_j\tlambda_l\,,\qquad k_l(z) = k_l + z\, \lambda_j\tlambda_l\,, 
\label{MomShift}
\end{eqnarray}
and on-shell conditions, here $k_j^2(z)=0=k_l^2(z)$, as well as overall
momentum conservation remain intact, due to $k_j(z)+k_l(z)=k_j+k_l$.  

Poles in the variable $z$ appear through propagators when intermediate
momenta go on-shell,
\begin{eqnarray}
\frac{i}{K^2_{\alpha}(z)}=\frac{i}{K_{\alpha}^2 +z \spba{l}.{K_{\alpha}}.{j}}  \,,  \label{Propagator}
\end{eqnarray} 
where $\alpha$ denoted a range of momentum vectors that contain momentum $k_l$ but not
$k_j$. The location of the pole is given by $ z_{\alpha} = - K_{\alpha}^2/
\spba{l}.{K_{\alpha}}.{j}$.
Near the singularity of the propagator~(\ref{Propagator}), the amplitude is
given by its factorization properties,
\begin{eqnarray}
\lim\limits_{z \rightarrow z_{\alpha}} A_n(z) = \sum_{h} A^h_{\alpha,L}(z_{\alpha}) {i
\over K_{\alpha}^2 + z \spba{l}.{K_{\alpha}}.{j}  } A^{-h}_{\alpha,R}(z_{\alpha})
\, ,
\label{TreeFactorization} \end{eqnarray} and uniquely defines the residue on
the pole in terms of the on-shell lower-point amplitudes $A^h_{\alpha,L}(z_{\alpha})$
and $A^{-h}_{\alpha,R}(z_{\alpha})$.  

By using Cauchy's residue theorem we can relate the amplitude at $z\!=\!0$,
$A_n(0)$, to its poles.  For a vanishing contour integral that encloses the
point $z=0$, we find,
\begin{eqnarray}
  0=\oint_{\cal C}\frac{dz}{2\pi i}
  \frac{A_n(z)}{z}\,\quad\rightarrow\quad A_n(0)=-\sum_{ {\rm
  poles}\,\, \alpha} {\rm Res}_{z=z_\alpha}\frac{A_n(z)}{z}.
\end{eqnarray}
The vanishing contour is associated to large complex momenta. In this region
power-counting arguments can be used to show the vanishing of the
function $A_n(z\rightarrow\infty)=0$. One may also include the residue at
$z=\infty$. However, this residue is not related to a factorization limit as
above (\ref{TreeFactorization}) and has to be obtained by other
means (see e.g.~\cite{Genhel}).

When there is no large-$z$ contribution, the physical amplitude $A_n(0)$ is
obtained explicitly by,
\begin{equation} 
	A_n(0) = \sum_{\alpha} \sum_{h} A^h_{\alpha,L}(z_{\alpha}) { i \over K^2_{\alpha} } A^{-h}_{\alpha,R}(z_{\alpha})  \,.  \label{BCFW} 
\end{equation}
The on-shell amplitudes with fewer legs, $A_{\alpha,L}^h$ and $A_{\alpha,R}^{-h}$, are evaluated in
kinematics that have been shifted by eq.~(\ref{SpinorShift}) with $k_i(z_{\alpha})$,
such that their intermediate momentum is on-shell,
$K_{\alpha}(z_{\alpha})^2=0\,$ where $z_\alpha = - K^2_{\alpha}/
\spba{l}.{K_\alpha}.{j}\,$.

The $n$-point amplitude $A_n$ is thus expressed in terms of sums over on-shell,
but complex continued, amplitudes with fewer legs.  These recursion relations
can be extended to massive QCD, more general other theories and beyond four
dimensional applications~\cite{Masses,GravityRecursion}.  Moreover, for certain
helicity configurations, they lead to new all-multiplicity expressions for
these amplitudes~\cite{Allorder}.  

\subsection{Recursion for Loop Amplitudes}
\label{LoopRecursion}

On-shell recursion have been generalized to loop-level in
refs.~\cite{Bootstrap,Genhel,CoeffRecursion}.  Following a similar strategy as
outlined in \sect{TreeRecursion} for tree amplitudes, loop amplitudes are analyzed
as functions of a complex parameter $z$.  When applying a
shift (\ref{SpinorShift}), we obtain,
\begin{equation} 
A_n^\oneloop(z) =  \Bigl[C_n(z) + R_n(z) \Bigr] \,.  \label{ComplexLoopAmpl}
\end{equation}
where $C_n(z)$ denotes the logarithmic part of the amplitudes, and $R_n(z)$ the
rational remainders (\ref{IntegralBasis}). In contrast to tree amplitudes, loop
amplitudes (\ref{ComplexLoopAmpl}) typically have branch cut singularities in
the variable $z$ originating from the logarithms within $C_n(z)$.

The way we deal with branch cut singularities is to focus on subparts of the
amplitude which are free of logarithms:
\begin{enumerate} 
\item the rational remainder: $R_n(z)$ 
\item the integral coefficients: $d^0_{i}(z)$, $c^0_{j}(z)$ or $b^0_{k}(z)$ within
	$C_n(z)$ 
\item $D$-dimensional coefficients: $e^0_h(z)$, $d^0_{i}(z)$, $c^0_{j}(z)$,
	$b^0_{k}(z)$ or $a^0_{l}(z)$. 
\end{enumerate}
These integral coefficients are introduced in sections~\ref{ExplicitTensorBasis}
and \ref{DExplicitTensorBasis}, respectively. These terms can be shown to
inherit factorization properties on physical poles from universal factorization
relations at loop-level. For an intermediate propagator going on-shell there are
typically three contributions for each internal state,
\begin{equation} 
\lim\limits_{K^2\rightarrow 0} A_n^\oneloop = A_L^\tree \, {i\over K^2 }
\,A_R^\oneloop 
+A_L^\oneloop \, {i\over K^2  } \, A_R^\tree
+A_L^\tree \, {i \, {\cal F}\over K^2} \,A_R^\tree\,.  
\label{FullOneLoopFact} 
\end{equation}
In the first two terms, one of the factorized amplitudes is a one-loop
amplitude and the other is a tree amplitude.  The last term will appear here
simply as a one-loop correction to the propagator.  (For details about
subtleties of the 'factorization function' ${\cal F}$ in massless theories we
refer to~\cite{BernChalmers}.)

When considering subparts of the full amplitude, additional singularities,
called spurious singularities, may appear. These singularities naturally cancel
out in the full amplitude. A method to compute contributions from spurious
singularities to rational terms $R_n$ will be discussed
in~\sect{SpuriousResidues}.

\subsection{Recursions for the Rational Part.}

The terms $R_n$ (\ref{IntegralBasis}), which are purely rational
in the spinor variables, cannot be computed using four-dimensional unitarity
methods.  On-shell recursion, however, allows us to construct these terms from
purely four-dimensional data.  Typically the information contained in the
recursion relation has to be completed with properties of the cut part.

\begin{figure}
\begin{center}
\includegraphics[width=.3\textwidth]{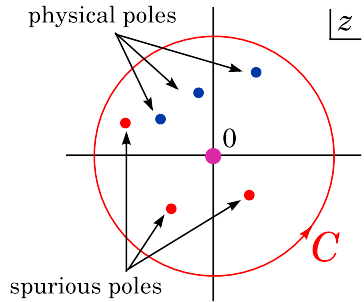}
 \caption{Using Cauchy's theorem, rational expressions in loop
amplitudes can be reconstructed from residues at poles in the 
complex plane.  The poles are of two types: physical and spurious.
All pole locations are known {\it a priori}.
Residues at physical poles are obtained from universal factorization
relations (\ref{FullOneLoopFact}).  Residues at spurious poles are obtained
from the cut parts.\\
{\it\tiny Reprinted \fig{LoopCauchyFigure} with permission from~\cite{BlackHatI} p.16. Copyright (2008) by the American Physical Society.}
} \label{LoopCauchyFigure} \end{center} \end{figure}

The central input for this approach is the detailed understanding of $R_n(z)$
as a function of $z$; we need to know the location of poles, their residues and
an integration contour, along which $R_n(z)$ vanishes.  Poles in the rational
terms $R_n(z)$ may be grouped into two classes as shown in
fig.~\ref{LoopCauchyFigure}: physical and spurious.  The physical poles are
present in the full amplitude $A_n(z)$, and correspond to genuine, physical
factorization singularities.  The spurious poles are not poles of $A_n(z)$ and
cancel between the cut parts $C_n(z)$ and rational remainders $R_n(z)$.  They
originate from the presence of tensor integrals in the underlying field-theory
representation of the amplitude, and appear as Gram determinant denominators
(\ref{DefGramDet}).  In unitarity approaches, inverse Gram determinants enter
integral coefficients through on-shell loop-momentum parametrization.  The
exponent of the inverse Gram determinant in integral coefficients can be
bounded by power-counting arguments.  These denominators give rise to spurious
singularities in individual terms.  

An example for the appearance of Gram determinants in unitarity approach can be
seen in the quadruple-cut computation in \sect{QuadrupleCutExample}. The
Gram determinant (see also~\cite{BlackHatI}),
\begin{eqnarray}
	\Delta(K_{12},k_3,k_4,k_5)=
	2s_{45}\spab4.{K_{12}}.5\spba4.{K_{12}}.5\,,
	\label{GramDet}
\end{eqnarray}
appears in the on-shell loop momenta (\ref{QuadCutMomenta}), through the factor
$1/\spba4.{K_{12}}.5$; the remaining invariants cancel. The second factor,
$1/\spba5.{K_{12}}.4$, appears in the parity conjugate on-shell momenta.  This
'chiral' dependence on inverse Gram determinants is typical for unitarity cuts
with massless corners.  In this particular example, the dependence on the Gram
determinant (\ref{GramDet}) cancels in the final form of the integral
coefficients in \eqn{QuadCutFinal}.  For the helicity configurations discussed
the integral coefficients are the same as in maximally supersymmetric
Yang-Mills and thus are effectively supersymmetric.  Supersymmetric
cancellations then reduce the naive power-count of loop momenta from four to
zero, such that no dependence on Gram determinant remains in the integral
coefficient. Explicit dependence of box-integrals on Gram determinant
singularities can be found for example in analytic expressions
in ref.~\cite{CutsQCD}.

Separating the different contributions of the shifted amplitude
(\ref{ComplexLoopAmpl}), we may write,
\begin{equation}
	R_n(z) = R^D_n(z) + R^S_n(z) + R^{{\rm large}\ z}_n(z)\,,
	\label{totrec} 
\end{equation}
where $R^D_n(z)$ contains all contributions from physical poles, $R^S_n(z)$ the
contributions from spurious poles, and $R^{{\rm large}\ z}_n(z)$ the possible
contributions from large deformation parameter $z$, if $R_n(z)$ does not vanish
there.  More explicitly, the rational terms can be expressed in terms of a
partial fraction decomposition in $z$,
\begin{eqnarray} 
&&R_n^D(z) = \sum_\alpha {A_\alpha \over z - z_\alpha}\,,
\qquad R^{{\rm large}\ z}_n(z) = \sum_{\sigma=0}^{\sigma_{\rm max}}
D_\sigma z^\sigma \,, \nonumber\\
&&R^S_n(z) = \sum_\beta \Biggl({B_\beta \over (z - z_\beta)^2} + {C_\beta \over
z - z_\beta } \Biggr)\,, 
\label{RnShiftGeneric} \end{eqnarray}
where the coefficients $A_\alpha, B_\beta, C_\beta, D_\sigma$ are functions of
the external momenta. The poles in $z$ in \eqn{RnShiftGeneric} are shown in
\fig{LoopCauchyFigure}.  The physical poles labeled by $\alpha$ are generically
single poles.
\footnote{Some shift choices may lead to double poles~\cite{DoublePole}; we can
generally avoid such shifts~\cite{Genhel}. A different approach based on
unitarity cuts was recently suggested~\cite{UnrealPole}.}
In general, in a renormalizable gauge theory, the spurious poles, labeled by
$\beta$, may be either single or double poles.  If $R_n(z)$ vanishes for large
$z$, the $D_\sigma$ are all zero.  If not, then $D_0$ gives a contribution to
the physical rational terms, $R_n(0)$.

\begin{figure}
\begin{center}
\centerline{
\epsfxsize 5 truein\epsfbox{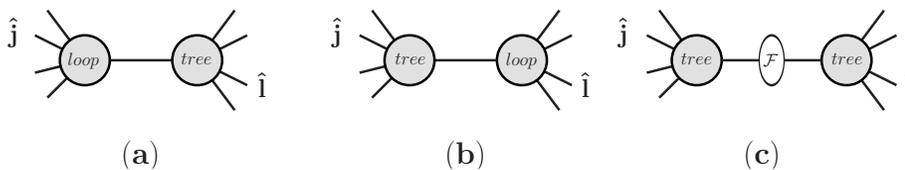}
}
 \caption{Diagrammatic contributions to on-shell recursion at one-loop for a
 $[j, l \rangle$ shift.  The labels '{\it loop}' and '{\it tree}' refer
 respectively to (lower-point) tree amplitudes $A^\tree$ and rational remainders of
 one-loop amplitudes $R$.  The central blob in (c) is the rational part of a
 one-loop factorization function ${\cal F}$~\cite{BernChalmers}.}
 \label{LoopGenericFigure} \end{center} \end{figure}

\subsubsection{Physical Residues.}
The contributions of the physical poles may be obtained efficiently using the
on-shell recursive terms represented by the diagrams in
\fig{LoopGenericFigure}.  The 'vertices' labeled by '{\it tree}' denote
tree-level on-shell amplitudes $A_m^\tree$, while the loop vertices '{\it
loop}' are the rational remainders of on-shell (lower-point) one-loop amplitudes
$R_m$, $m<n$, as defined in \eqn{CutRational}.  The contribution in
\fig{LoopGenericFigure}(c) involves the rational part of the additional
factorization function ${\cal F}$~\cite{BernChalmers}.  It only appears in
multi-particle channels, and only if the tree amplitude contains a pole in that
channel.  Each diagram is associated with a physical pole (\ref{Propagator}) in
the $z$ plane, as in \fig{LoopCauchyFigure},  and is computed similarly to the
recursive diagrams at tree-level (\ref{BCFW}). Further details for the
computation of the recursive diagrams in \fig{LoopGenericFigure} has been
described in refs.~\cite{Bootstrap,OneLoopMHV,OnShellReview}.

\subsubsection{Spurious Residues.}
\label{SpuriousResidues}
One approach to compute spurious residues is to use a completion of the
integral functions, so called 'cut-completion'~\cite{Bootstrap,Genhel}. To this
end, the integral basis is adjusted to subtract off spurious poles within their
integral coefficients.  This in turn moves all spurious poles from the rational
part $R_n$ in (\ref{ComplexLoopAmpl}) to the cut part $C_n$, such that the
redefined rational remainder has poles only at the position of physical factorization
singularities.  The coefficients $B_\beta$ and $C_\beta$ vanish in this
approach.  Attention has to be paid to the factorization equation which will
mix cut part and rational part non-trivially.  This approach has led to the
computation of the rational terms for a variety of one-loop MHV amplitudes with
an arbitrary number of external legs~\cite{Bootstrap,Genhel}, as well as for
six-point amplitudes.  

For the purposes of a numerical program, however, it is simpler to
extract the spurious residues from the known cut parts~\cite{BlackHatI}.  These
residues, being absent in the full loop amplitude, are guaranteed to be the
negatives of the spurious-pole residues in the rational remainder.  That is,
the spurious contributions are,
\begin{equation}
\lim\limits_{z\rightarrow z_\beta} C_n(z)=-\Biggl({B_\beta \over (z -
z_\beta)^2} + {C_\beta \over z - z_\beta } \Biggr)+\Ord(z)\,, \label{spureqn}
\end{equation}
where $C_n(z)$ is the shifted cut part appearing in \eqn{CutRational}. Terms
containing logarithms in the kinematic invariants cancel, but may as well be
ignored for simplicity. The spurious poles at $z_\beta$ correspond to the vanishing
of shifted Gram determinants, $\Delta_m(z_\beta)=0$ for $m=3,4$, associated
with triangle and box integrals.

In order to compute the rational parts $R_n$ we need to extract all residues
$B_\beta$ and $C_\beta$ from the cut pieces $C_n(z)$.  For this we evaluate the
integral coefficients $d_i(z),c_j(z)$ and $b_k(z)$ numerically for complex, shifted momenta in
the vicinity of the spurious pole.  We also need to evaluate the loop
integrals.  This is done after using as input analytic series expansions of the
integrals around vanishing Gram determinants,
\begin{eqnarray}
\I_i(z)&\inlimit^{z\rightarrow z_\beta}& \I_i^\beta(z)+\mbox{logarithms}\,\nonumber\\
\I_i^\beta(z)&\,\,=&\!\!(z-z_\beta)^k \Bigl(\rho_1+ (z-z_\beta)\rho_2+\ldots \Bigr)\,.
\label{IntegralExpansion}
\end{eqnarray}
The coefficients $\rho_l$ are rational functions in the kinematic variables
which take a universal form.  Examples of this procedure may be found
in~\cite{BlackHatI} with further details for generating these expansions using
a dimension-shifting formula~\cite{BDKIntegrals}  may be found
in~\cite{BFReview}. It is important to have a precomputed analytic expansion
available, as the logarithmic terms have to be dropped by hand; these terms do
not cancel spurious poles in rational terms, but rather cancel between
different integral functions.  Thus we may avoid computing any logarithms or
polylogarithms at complex momentum values.  The expression obtained by
replacing $C_n(z)$ according to these rules, in the vicinity of $z_\beta$, will
be denoted by $C_n^\beta(z)$.

\subsubsection{Discrete Fourier Sum for Spurious Residues.}

\begin{figure}
\begin{center}
\includegraphics*[width=0.4\textwidth]{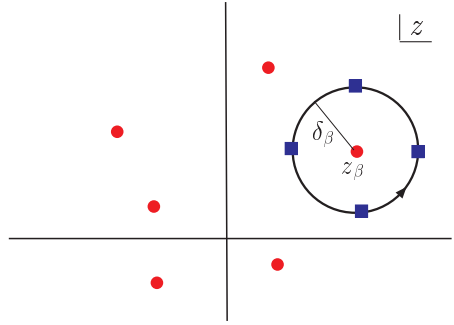}
 \caption{We obtain the residue at the spurious pole located at $z = z_\beta$
 in the complex $z$ plane by a discrete Fourier sum, evaluating $C_n^\beta(z)$
 on the (blue) squares on the circle of radius $\delta_\beta$ centered on
 $z_\beta$.  The locations of other poles are represented by (red) dots.  We
 ensure that $\delta_\beta$ is sufficiently small so that other poles give a
 negligible contribution to the residue.\\
{\it\tiny Reprinted \fig{ContourRatFigure} with permission from~\cite{BlackHatI} p.22. Copyright (2008) by the American Physical Society.}
 } \label{ContourRatFigure} \end{center}
 \end{figure}


The spurious-pole residues can be extracted from the cut parts
by using a discrete Fourier sum.  We evaluate $C_n^\beta(z)$ at 
$m$ points equally spaced around a circle of radius $\delta_\beta$ 
in the $z$ plane, centered on the pole location $z_\beta$, 
as depicted in \fig{ContourRatFigure}; {\it i.e.},
$z = z_\beta + \delta_\beta e^{2\pi ij/m}$, for $j=1,2,\ldots,m$.

We can extract the coefficients $B_\beta$ and 
$C_\beta$ in \eqn{RnShiftGeneric} via,
\begin{eqnarray}
B_\beta &\simeq& -{1\over m} \sum_{j = 1}^{m} 
          \Bigl[ \delta_\beta  \, e^{2 \pi ij /m} \Bigr]^2
          C_n^\beta(z_\beta+\delta_\beta e^{2 \pi i j/m})\,, \nonumber \\
C_\beta &\simeq& -{1\over m} 
        \sum_{j = 1}^{m} \delta_\beta \, e^{2 \pi ij /m} \,
          C_n^\beta(z_\beta+\delta_\beta e^{2 \pi i j/m})\,.
\end{eqnarray}
Both $m$, the number of evaluations, and $\delta_\beta$ are adjusted to
optimize efficiency and numerical precisions.
In general, an increase in $m$ increases the precision, but at the cost of
computation time. We choose $\delta_\beta$ to be much smaller than the distance
to nearby poles, but not so small as to lose numerical precision.


\subsection{Recursion Relations for Integral Coefficients}\label{rerelfintco}
\label{IntegralCoeffRecursionSubsection}

Here we address the question of whether we can also apply recursion relations
to the cut containing pieces $C_n(z)$.  In fact it has been shown
in~\cite{CoeffRecursion} that it is possible to construct recursions, not for
the full amplitudes, but in certain cases for the rational coefficients of
the integrals.  Such recursions can have important implication for the
computation of rational terms, as we will discuss below in \sect{SpuriousRecursions}.

With unitarity methods available, one of the central motivations for coefficient
recursion are efficiency gains and insights for the recursive computation of
rational terms. Rational recursion relies on the data from cut containing terms;
the residues of Gram determinant poles.  Gram determinant singularities are
tied to their associated integral coefficients.  A direct recursive approach to
compute integral coefficients directly gives access to spurious residues of Gram
determinant poles.  

A firm grasp of the analytic properties of the integral coefficients is
required. These properties differ from those of full amplitudes; in general
these coefficients, being subparts of the full amplitude, contain physical
poles as well as spurious poles of their own. Very much like for the case of
rational terms, spurious poles originate from reduction of higher point tensor
integrals and are associated to Gram determinant denominators. 

The spurious poles are harder to deal with in a purely recursive way.  An
intricate web of cancellations between distinct integral coefficients guided by
the integral functions leads to the cancellation of spurious poles between
various logarithmic terms. Here, we do not want to disentangle the implicit
consistency conditions.  Rather, with an understanding of which spurious poles
appear in which integral coefficient, we try to maneuver around them.  A detailed
discussion of the factorisation properties of one-loop amplitudes, as well the
spurious singularities that appear, may be found in
refs.~\cite{UnitarityMethod,BernChalmers}.  

\begin{figure}
\begin{center}
\centerline{
\epsfxsize 2.5 truein\epsfbox{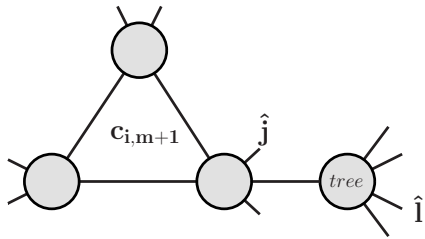}
}
 \caption{Diagrammatic contributions to on-shell recursion for a three-point
 integral coefficient $c_{i,n}(z)$ for a $[j, l \rangle$ shift.  The
 left-hand-side indicates a triangle loop diagram that gives rise to the input
 lower-point coefficient $c_{i,m+1}$.  The recursion is built from sewing a
 lower-point integral coefficient $c_{i,m+1}(z_\alpha)$ and a tree amplitude $A_{n-m+1}(z_\alpha)$
 The vertices of the triangle graph are shown as the tree amplitudes of the
 underlying triple cut.  The shifted legs are picked from within the same
 cluster of legs.  } \label{CoefficientRecFigure} \end{center} \end{figure}

The factorisation of amplitudes follows from the combined behaviour of integral
functions and the integral coefficients in the factorisation limit.  If we turn
this around, given the general factorisation (\ref{FullOneLoopFact}) of an
amplitude and given factorization properties of integral
functions~\cite{UnitarityMethod,BernChalmers}, we may then determine the
factorisation properties of the integral coefficients.  By applying this logic
to multi-particle factorisations (\ref{FullOneLoopFact}) we conclude that the
coefficients behave as if they were tree amplitudes as long as the
factorisations are entirely within a cluster of legs (and are not on the
momentum invariant of the entire cluster).  That is the coefficient behaves as,
\begin{eqnarray}
c_{i,n} \ \inlimit^{K^2 \rightarrow 0} \ \ \sum_h
A^h_{n-m+1}\,\frac{i}{K^2}\,c_{i,m+1}^{-h}\,,
\end{eqnarray}
with notation as indicated in \fig{CoefficientRecFigure}. 
For convenience, from now on we will use the notation $c_i$ for all
integral coefficients; coefficients of two-point functions, three-point
functions etc.

Assuming that the spurious denominators do not pick up a $z$ dependence ---
below we describe simple criteria for ensuring this  --- we  obtain a recursion
relation for the coefficients which strikingly is no more complicated than for
tree amplitudes,
\begin{eqnarray}
c_{i,n}(0) \; = \; \sum_{\alpha,h}  {A^h_{n-m_\alpha+1}(z_\alpha) \, {i\over
K^2_{\alpha}}\, c^{-h}_{i,m_\alpha+1}(z_\alpha)} \,,
\label{CoeffRecur} \end{eqnarray}
where $A^h_{n-m_\alpha +1}(z_\alpha)$ and $c^h_{i,m_\alpha+1}(z_\alpha)$ are
shifted tree amplitudes and coefficients evaluated at the residue value
$z_\alpha$\,, $h$ denotes the helicity of the intermediate state corresponding
to the propagator term $i/K^2_{\alpha}$\,. In this expression one should only
sum over a limited set of poles; if the shifts are chosen from within a
cluster, the only poles that should be included are from within the kinematic
invariants formed from the momenta making up the cluster. Pictorially, this
coefficient recursion relation is shown in \fig{CoefficientRecFigure}.

Several consistency requirements have to be fulfilled for on-shell
recursions of integral coefficients.  Simple criteria for a valid recursion
were given in~\cite{CoeffRecursion}: 
\begin{enumerate}
\item The shifted tree amplitude, on the side of the cluster undergoing
	recursion, vanishes for large $z$.
\item The loop-momentum dependent kinematic poles are unmodified by the shift
	(\ref{MomShift}).
\end{enumerate}
In addition to the standard requirement, that the shifted coefficient vanishes
for large $z$, we have criteria (ii) which sidesteps contributions from
spurious poles.  In particular, this relies on the assumption that spurious
poles in $z$ appear whenever loop-momentum dependent propagator give rise to
non-zero residues. 

The above criteria are fulfilled for selecting a shift within the same cluster
of legs in addition to a requirement on the helicity structure of the
respective cluster.  For a particular set of helicity amplitudes, so called split
helicity amplitudes, coefficient recursions take a very simple form. (Split
helicity refers to color-ordered amplitudes with all like helicities adjacent;
$A_n(-\ldots-,+\ldots+)$.) For these amplitudes computations were performed for
all multiplicity computations in supersymmetric Yang-Mills, pure QCD as well as
all 7-point amplitudes for $W/Z +3$-jet
production~\cite{PRLW3BH,W3jDistributions,TeVZ}, including one as well as two
fermion lines.  Further examples include integral coefficients with
split-helicity corners.

\subsection{Spurious Recursions.}
\label{SpuriousRecursions}

When recursions for integral
coefficients can be setup, we can formulate analogous recursions for rational terms.
These 'auxiliary recursions' then directly give access to spurious residues within
rational terms.  We obtain in this way a purely recursive approach for
rational terms, where no reference to integral expansions of cut pieces is needed.  

\begin{figure}
\begin{center}
\centerline{
\epsfxsize 4.0 truein\epsfbox{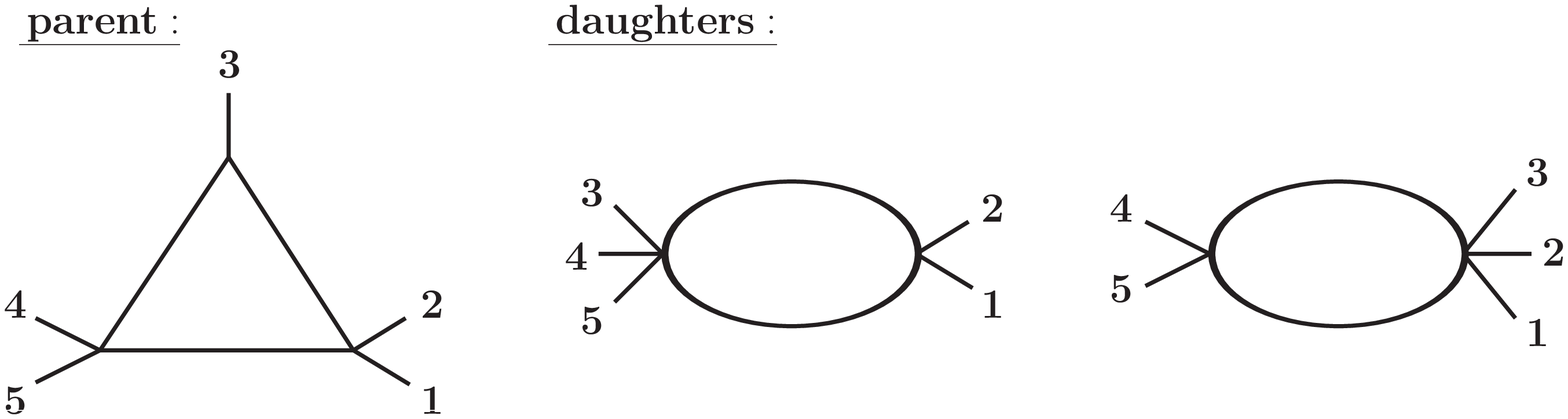}
}
 \caption{
 Parent and daughter integral functions. The Gram determinant singularity
 associated to the parent 3-point integral is $\Delta(K_{12},k_3,K_{45})$. It
 is present in its integral coefficient as well as in the coefficients of both
 daughter 2-point integrals.}
 \label{GramDetDaugher} \end{center} \end{figure}

In fact, given that we shift integral coefficients within one of its corners,
integral functions stay inert under the shift; they only depend on un-shifted
momenta. Similarly, we may expand integrals around their Gram-determinant poles
(see ~\eqn{IntegralExpansion}), still yielding terms independent of the
shift-variable. Similar steps can be followed for integral coefficients of
lower-point (daughter) integrals which inherit a given Gram determinant
singularity. The structure of parent and daughter integral coefficients is
indicated in \fig{GramDetDaugher}.  

Putting all pieces together, we obtain an on-shell recursion relation for
rational terms, which is valid close to a given vanishing Gram determinant,
\begin{eqnarray}
\lim\limits_{\Delta_\beta\rightarrow 0}R_n&\sim&-\sum_i \I_i^\beta
c_{i,n}^\beta = -\sum_{i,\alpha,h} {A^h_{n-m_\alpha+1}(z_\alpha) \, {i\over
K^2_{\alpha}}\, \I_i^\beta c^{-h}_{i,m_\alpha+1}(z_\alpha)}\,\nonumber\\ &=&
\sum_{\alpha,h} {A^h_{n-m_\alpha+1}(z_\alpha) \, {i\over K^2_{\alpha}}\,
R^{-h}_{m_\alpha+1}(z_\alpha)}\,,
\label{SpuriousRational} \end{eqnarray}
where $\Delta_\beta$ denotes the Gram determinant around which we intend to
expand the rational term. $\I_i^\beta$ denote the integral functions, which are
expanded around the location of the zero of the Gram determinant as
in~\eqn{IntegralExpansion}. In addition, only their rational terms are kept.
The symbol $\alpha$ labels the physical poles that need to be considered in the
coefficient recursions of the parent integral coefficient.  For convenience
integral coefficients are denoted by $c_i$, for any of the integral functions
somewhat differing from our earlier notation in \sect{UnitaritySection}.

We thus obtain a recursion relation for the rational terms, which is exact near
the Gram determinant singularities. Contour integrals whether continuous or
discrete can then be used to extract the exact value of the spurious residues
of~\eqn{RnShiftGeneric}. 

Thus obtained recursive approaches allow to compute rational terms from purely
recursive methods, albeit, using a set of well chosen auxiliary shifts.  These
methods have already been applied for the computations of split helicity
scattering amplitudes, needed in the computation of \Wjjj-jet and \Zjjj-jet
production~\cite{PRLW3BH,W3jDistributions,TeVZ} leading to greatly improved
computation time for these pieces. A much wider class of computations seems in
reach using the above newly developed recursive methods.


\extraskip
\section{Conclusions}
\label{ConclusionSection}

The computations of observable cross-sections rely on the composition of
several components, which are linked through the fundamental QCD Lagrangian.
Here we described modern approaches for evaluating the hard scattering part at
next-to-leading order, which can be programmed.  The value of a first principle
understanding of scattering processes in addition to the increased quantitative
control motivates the quest for cross sections at NLO.  

The description of processes with complex final states at NLO is one of the
central achievements of many recent developments of numerical
on-shell~\cite{BCFW,Bootstrap,CoeffRecursion,BlackHatI} and unitarity
methods~\cite{UnitarityMethod,DdimUnitarity,BCFUnitarity,OPP,BlackHatI,GKM}.
These methods are already used by a new generation of
tools~\cite{CutTools,BlackHatI,GZ,W3EGKMZ,OtherLargeN,GKW,SAMURAI,Madloop,BBU}.
Such numerical methods scale very well as the number of external states
increases and have already lead to a new understanding for several proton
scattering processes at hadron
colliders~\cite{MStop,OPPNLO,W4jets,OtherUnitarityNLO,Madloop} including
$W/Z+3$-jet and $W+4$-jet production~\cite{PRLW3BH,EMZW3Tev,W3jDistributions,W4jets}. Being
key backgrounds to many new physics signals including supersymmetry searches,
the explicit results emphasizes the importance of modern field-theory methods
for precision prediction of hadron collider physics. 

Numerical on-shell and unitarity methods allow us to automate computations of
scattering amplitudes in a numerically stable and efficient way.  These methods
exploit, for example, the discontinuities across branch cuts to construct
amplitudes.  Efficiency and stability originates then in two facts: Firstly,
discontinuities are expressed in terms of purely on-shell information, namely,
on-shell tree-amplitudes. Evaluation of on-shell trees is fast and numerically
stable.  In addition, this allows us to ignore ghosts and gauge fixing
altogether and reduces the use of redundant unphysical information.  

These modern ideas presented here my also help with issues beyond the
computation of matrix elements. New insights may impact on various other
components of multi-jet computations.  This includes subtraction methods for
integrating parton level NLO computations, showering and formulation of new
observables.

The central strategy, leading to the success and popularity of these
developments, is to make maximal use of physical principles and mathematical
structures in order to obtain efficient and robust phenomenology.  This kind of
work, is at the cross-roads of theory and phenomenology has led to many recent
insights, on the more formal
side~\cite{BDS,Gravity,BlackHatI,GKM,NimaAllLoop,ABCT} as well as of more
phenomenological nature~\cite{Wpol}.


\section*{Acknowledgments}

I am grateful to Giovanni Diana, Lance Dixon, Stefan Hoeche, Darren Forde, David
Kosower, Daniel Ma\^{\i}tre for related collaborations and in particular to Zvi
Bern, Fernando Febres Cordero and Kemal Ozeren for many stimulating discussions
on the topics described here.  This work is supported by a fellowship from the
US LHC Theory Initiative through NSF grant PHY-0705682.  I am also grateful for
the hospitality of the KITP during the very stimulating workshop {\it Harmony
of Scattering Amplitudes}, where parts of this review were completed.  This
research was supported in part by the National Science Foundation under Grant
No. NSF PHY05-51164.


\end{document}